\documentclass[final,5p,times,twocolumn]{elsarticle} 

\usepackage{graphicx}

\usepackage{amssymb}

\usepackage{amsfonts}
\usepackage{amsmath}
\usepackage{upgreek} 

\usepackage{bm} 

\usepackage{lineno}

\usepackage[dvips]{color}

\journal{Astroparticle Physics}

\begin{document}

\begin{frontmatter}



\title{KASCADE-Grande measurements of energy spectra for elemental groups of cosmic rays}


\author[address:1]{W.D.~Apel}
\author[address:2]{J.C.~Arteaga-Vel\'azquez}
\author[address:1]{K.~Bekk}
\author[address:3]{M.~Bertaina}
\author[address:1,address:4]{J.~Bl\"umer}
\author[address:1]{H.~Bozdog}
\author[address:5]{I.M.~Brancus}
\author[address:3,address:6]{E.~Cantoni\fnref{address:c}}
\author[address:3]{A.~Chiavassa}
\author[address:4]{F.~Cossavella\fnref{address:a}}
\author[address:1]{K.~Daumiller}
\author[address:7]{V.~de Souza}
\author[address:3]{F.~Di~Pierro}
\author[address:1]{P.~Doll}
\author[address:1]{R.~Engel}
\author[address:1]{J.~Engler}
\author[address:4]{M. Finger}
\author[address:4]{B.~Fuchs}
\author[address:8]{D.~Fuhrmann\corref{corresponding_author}}
\ead{fuhrmann@uni-wuppertal.de}
\author[address:1]{H.J.~Gils}
\author[address:8]{R.~Glasstetter}
\author[address:9]{C.~Grupen}
\author[address:1]{A.~Haungs}
\author[address:1]{D.~Heck}
\author[address:10]{J.R.~H\"orandel}
\author[address:4]{D.~Huber}
\author[address:1]{T.~Huege}
\author[address:8]{K.-H.~Kampert}
\author[address:4]{D.~Kang}
\author[address:1]{H.O.~Klages}
\author[address:4]{K.~Link}
\author[address:11]{P.~{\L}uczak}
\author[address:4]{M.~Ludwig}
\author[address:1]{H.J.~Mathes}
\author[address:1]{H.J.~Mayer}
\author[address:4]{M.~Melissas}
\author[address:1]{J.~Milke}
\author[address:5]{B.~Mitrica}
\author[address:6]{C.~Morello}
\author[address:1]{J.~Oehlschl\"ager}
\author[address:1]{S.~Ostapchenko\fnref{address:b}}
\author[address:4]{N.~Palmieri}
\author[address:5]{M.~Petcu}
\author[address:1]{T.~Pierog}
\author[address:1]{H.~Rebel}
\author[address:1]{M.~Roth}
\author[address:1]{H.~Schieler}
\author[address:4]{S.~Schoo}
\author[address:1]{F.G.~Schr\"oder}
\author[address:12]{O.~Sima}
\author[address:5]{G.~Toma}
\author[address:6]{G.C.~Trinchero}
\author[address:1]{H.~Ulrich}
\author[address:1]{A.~Weindl}
\author[address:1]{J.~Wochele}
\author[address:1]{M.~Wommer}
\author[address:11]{J.~Zabierowski}

\cortext[corresponding_author]{Corresponding author. Address: Fachbereich Physik, Universit\"at Wuppertal, Gau\ss str. 20, 42119 Wuppertal, Germany. Tel.: +49 202 4392640; fax: +49 202 4392662.}
\fntext[address:a]{Present address: Max-Planck-Institut Physik, M\"unchen, Germany}
\fntext[address:b]{Present address: University of Trondheim, Norway}
\fntext[address:c]{Present address: Istituto Nazionale di Ricerca Metrologica, Torino, Italy}

\address[address:1]{Institut f\"ur Kernphysik, KIT - Karlsruher Institut f\"ur Technologie, Germany}
\address[address:2]{Universidad Michoacana, Instituto de F\'{\i}sica y Matem\'aticas, Morelia, Mexico}
\address[address:3]{Dipartimento di Fisica, Universit\`a degli Studi di Torino, Italy}
\address[address:4]{Institut f\"ur Experimentelle Kernphysik, KIT - Karlsruher Institut f\"ur Technologie, Germany}
\address[address:5]{National Institute of Physics and Nuclear Engineering, Bucharest, Romania}
\address[address:6]{Osservatorio Astrofisico di Torino, INAF Torino, Italy}
\address[address:7]{Universidade S$\tilde{a}$o Paulo, Instituto de F\'{\i}sica de S\~ao Carlos, Brasil}
\address[address:8]{Fachbereich Physik, Universit\"at Wuppertal, Germany}
\address[address:9]{Department of Physics, Siegen University, Germany}
\address[address:10]{Dept. of Astrophysics, Radboud University Nijmegen, The Netherlands}
\address[address:11]{National Centre for Nuclear Research, Department of Cosmic Ray Physics, Lodz, Poland}
\address[address:12]{Department of Physics, University of Bucharest, Bucharest, Romania}

\begin{abstract}
The KASCADE-Grande air shower experiment \cite{lit:kascade_grande_allgemeim_nimpaper} consists of, among others, a large scintillator array for measurements of charged particles, $N_{\mathrm{ch}}$, and of an array of shielded scintillation counters used for muon counting, $N_{\upmu}$. 
KASCADE-Grande is optimized for cosmic ray measurements in the energy range 10~PeV to about 2000~PeV, where exploring the composition is of fundamental importance for understanding the transition from galactic to extragalactic origin of cosmic rays.
Following earlier studies of the all-particle and the elemental spectra reconstructed in the knee energy range from KASCADE data \cite{lit:kascade-unfolding}, we have now extended these measurements to beyond 200~PeV. 
By analysing the two-dimensional shower size spectrum $N_{\mathrm{ch}}$ vs. $N_{\upmu}$ for nearly vertical events, we reconstruct the energy spectra of different mass groups by means of unfolding methods over an energy range where the detector is fully efficient. The procedure and its results, which are derived based on the hadronic interaction model \mbox{QGSJET-II-02} and which yield a strong indication for a dominance of heavy mass groups in the covered energy range and for a knee-like structure in the iron spectrum at around 80~PeV, are presented. 
This confirms and further refines the results obtained by other analyses of KASCADE-Grande data, which already gave evidence for a knee-like structure in the heavy component of cosmic rays at about 80~PeV \cite{lit:heavy_knee_paper}.
\end{abstract}


\end{frontmatter}


\section{Introduction}
\label{Sec:Introduction}
The spectrum of cosmic rays follows roughly a power law behaviour ($\propto E^{\gamma}$, with $\gamma \approx -2.7\ldots-3.3$) over many orders of magnitude in energy, overall appearing rather featureless. However, there are a few structures observable. In 1958, Kulikov and Khristiansen \cite{lit:first_knee_measurement} discovered a distinct steepening in the electron size spectrum measured for extensive air showers initiated by cosmic rays, corresponding to a change of the power law slope of the all-particle energy spectrum at few PeV. Three years later, Peters \cite{lit:peters} concluded that the position of this kink, also called the ``knee'' of the cosmic ray spectrum, will depend on the atomic number of the cosmic ray particles if their acceleration is correlated to magnetic fields. This would mean that the spectra of lighter and heavier cosmic ray mass groups exhibit knee structures with growing energy successively. About half a century later, EAS-TOP observations \cite{lit:EAStop_knee1,lit:EAStop_knee2} and, in a more detailed analysis, the KASCADE experiment \cite{lit:kascade_allgemein_nimpaper,lit:kascade-unfolding} showed that the change of spectral index detected by Kulikov and Khristiansen could be caused by a decrease of the so far quantitatively dominating light component of cosmic rays. More precisely, the KASCADE results \cite{lit:kascade-unfolding} have proved that the knee in the all-particle spectrum at about 5~PeV corresponds to a decrease of flux observed for light cosmic ray primaries, only. This result was achieved by means of an unfolding analysis disentangling the convoluted energy spectra of five mass groups from the measured two-dimensional shower size distribution of electrons and muons at observation level. 

There are numerous theories about details of the origin, acceleration, and propagation of cosmic rays. Concerning the knee positions of individual primaries, some of the models predict, in contrast to the magnetic rigidity dependence considered by Peters \cite{lit:peters}, a correlation with the mass of the particles (e.g. cannonball model \cite{lit:cannonball}). Hence, it is of great interest to verify whether also the spectra of heavy cosmic ray mass groups exhibit analogous structures and if so, at what energies. The KASCADE-Grande experiment \cite{lit:kascade_grande_allgemeim_nimpaper}, located at Karlsruhe Institute of Technology (KIT), Germany, extends the accessible energy range of KASCADE to higher energies up to around 2000~PeV, and allows by this to investigate the cosmic ray energy spectra and composition at regions where the iron knee is expected.

The determination of the energy where the iron knee occurs enables the validation of the various theoretical models. Following this purpose, the KASCADE-Grande measurements have been analysed with straightforward but robust analysis methods yielding an evidence for a steepening in the cosmic ray all-particle spectrum at about 80~PeV \cite{lit:KG_all_part_spec_paper}, which corresponds to a knee-like structure in the heavy component of cosmic rays at about this energy \cite{lit:heavy_knee_paper}. In order to verify and to refine the obtained results, an unfolding technique has been used similar to the one applied to KASCADE data \cite{lit:kascade-unfolding, lit:KASCADE_low_energy_interaction_model}, but now for the KASCADE-Grande energy range and based on the interaction models \mbox{QGSJET-II-02}~\citep{lit:qgsjet-ii-model-zitat1, lit:qgsjet-ii-model-zitat2} and \mbox{FLUKA} 2002.4~\citep{lit:FlukaAllgemeinZitat1,lit:Fluka2002_4_Zitat1,lit:FlukaAllgemeinZitat2}. The unfolding method used will be outlined, and the results, which yield a strong indication for a dominance of the cosmic ray all-particle spectrum by heavy mass groups in the observed energy range and for a knee in the iron spectrum at about 80~PeV, will be presented in this publication. A more detailed description of the unfolding analysis can be found in \cite{lit:phd_fuhrmann}. 
\section{Outline of the analysis}
\label{Sec:outline_analysis}
\subsection{Data}
\label{Sec:data}
The KASCADE-Grande experiment\footnote{Located at Karlsruhe Institute of Technology (KIT), $49.1^\circ$~N, $8.4^\circ$~E. Observation level 110~m a.s.l., corresponding to an average atmospheric depth of $1022\:\mathrm g/ \mathrm{cm}^2$.} measures air showers initiated by primary cosmic rays in the energy range\footnote{In this work, the upper energy is limited to about 200~PeV since data statistics are too small in the used sample of vertical showers at higher energies.} 10~PeV to about 2000~PeV. It consists of a large scintillator array for measurements of charged particles, $N_{\mathrm{ch}}$, and of an array of shielded scintillation counters used for muon counting, $N_{\upmu}$, with a resolution of $\lesssim 15\%$ and $\lesssim 20\%$, respectively. A comprehensive description of the experiment, the data acquisition and the event reconstruction, as well as the achieved experimental resolutions is given in \cite{lit:kascade_grande_allgemeim_nimpaper,lit:KG_all_part_spec_paper,lit:phd_fuhrmann}. 

\begin{figure}
 \centering
 \includegraphics[width=0.985\columnwidth,bb= 0 0 567 459]{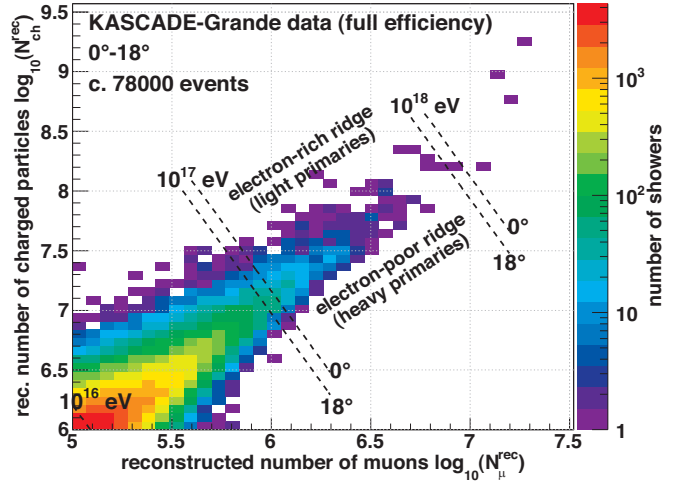}
  \caption{Two-dimensional distribution of the shower sizes (total number of charged particles and of muons) measured with KASCADE-Grande and used for this analysis. Diverse quality cuts are applied. Furthermore, only events with zenith angles $\theta \leq 18^\circ$ and with shower sizes for which the experiment is fully efficient (above $\log_{10}(N_{\mathrm{ch}}^{\mathrm{rec}})\approx6.0$ and $\log_{10}(N_{\upmu}^{\mathrm{rec}})\approx5.0$) are considered. In addition, a roughly estimated energy scale is indicated. Since KASCADE-Grande measures the shower sizes at atmospheric depths beyond the shower maximum, electron-rich showers are initiated preferentially by light primaries, and electron-poor showers by heavy ones, respectively (this is indicated in the figure, too).}
 \label{fig:measured_shower_size_plane}
\end{figure}

In Fig.~\ref{fig:measured_shower_size_plane}, the two-dimensional shower size spectrum number of charged particles $\log_{10}(N_{\mathrm{ch}}^{\mathrm{rec}})$ vs. number of muons $\log_{10}(N_{\upmu}^{\mathrm{rec}})$ measured with KASCADE-Grande, and used as basis for this analysis, is depicted. 
Only events with shower sizes for which the experiment is fully efficient are considered, i.e. $\log_{10}(N_{\mathrm{ch}}^{\mathrm{rec}})\gtrsim6.0$ and $\log_{10}(N_{\upmu}^{\mathrm{rec}})\gtrsim5.0$. In order to avoid effects due to the varying attenuation of the shower sizes for different angles of incidence, the data set used is restricted to showers with zenith angles $\theta \leq 18^\circ$. Furthermore, a couple of quality cuts are applied (cf. \cite{lit:kascade_grande_allgemeim_nimpaper,lit:KG_all_part_spec_paper,lit:phd_fuhrmann}). Finally, the measurement time covers approximately $1\,318$~days resulting in c. $78\, 000$ air shower events having passed all quality cuts and contributing to Fig.~\ref{fig:measured_shower_size_plane}, and yielding an exposure of $164\, 709$~m$^2$~sr~yr.
\subsection{Analysis}
\label{Sec:analysis}
The analysis' objective is to compute the primary energy spectra of $N_\mathrm{nucl}=5$ cosmic ray mass groups\footnote{This number of considered primaries has been found to yield a good compromise between the minimum number of primaries needed to describe the measured data sufficiently well, and the dispersion effects due to the limited resolution of the shower sizes (cf. Section~\ref{Sec:reponse_matrix} for details).}, represented by protons (p), as well as helium (He), carbon (C), silicon (Si), and iron (Fe) nuclei. 

The convolution of these sought-after differential fluxes $\mathrm d J_n/\mathrm d\, \mathrm{log}_{10}E$ of the primary cosmic ray nuclei $n$, with $n=1\ldots N_\mathrm{nucl}$, having an energy $E$ into the measured number of showers $N_i$ that is contributing to the content of the specific charged particle and muon number bin $\left( \log_{10}(N_{\mathrm{ch}}^{\mathrm{rec}}), \log_{10}(N_{\upmu}^{\mathrm{rec}}) \right)_{i}$ in Fig.~\ref{fig:measured_shower_size_plane}, can be described by an integral equation:
\begin{equation}\label{eq:faltung_des_primaerspektrums}
\begin{split}
N_i&=2\uppi A_{\mathrm f} T_{\mathrm m} \sum_{n=1}^{N_\mathrm{nucl}} \intop_{0^\circ}^{18^\circ} \intop_{-\infty}^{+\infty} \frac{\mathrm d J_n}{\mathrm d\, \mathrm{log}_{10}E}\; p_n\; \sin\theta\ \cos\theta\ \mathrm d\,\mathrm{log}_{10}E\  \mathrm d\theta   \; \; \; ,  \\
&\mathrm{with} \; \; \; p_n=p_n\left(   \left(  \mathrm{log}_{10}N_{\mathrm{ch}}^{\mathrm{rec}}, \mathrm{log}_{10}N_{\upmu}^{\mathrm{rec}}   \right)_i \ | \    \mathrm{log}_{10}E    \right)  \; \; \; .
\end{split}
\end{equation}
The sampling area $A_{\mathrm f}$ and the measurement time $T_{\mathrm m}$ are constants. The factor $2\uppi$ accounts for the integration over the azimuth angle, of which the data do not show any significant dependence. Hence, the integration over the whole solid angle has been reduced to one over the zenith angle range $0^\circ \leq \theta \leq 18^\circ$.

The conditional probabilities $p_n$ are originating from a convolution merging the intrinsic shower fluctuations $s_n$, the trigger and reconstruction efficiency $\varepsilon_n$, as well as the reconstruction resolution and systematic reconstruction effects $r_n$:
\begin{equation}\label{eq:pn}
\begin{split}
p_n\left(   \left(  \mathrm{log}_{10}N_{\mathrm{ch}}^{\mathrm{rec}}, \mathrm{log}_{10}N_{\upmu}^{\mathrm{rec}}   \right)_i \ | \    \mathrm{log}_{10}E    \right)=\\
\intop_{-\infty}^{+\infty} \intop_{-\infty}^{+\infty} s_n\ \varepsilon_n \ r_n \ \mathrm d \,\mathrm{log}_{10}N_{\mathrm{ch}}^{\mathrm{tru}} \ \mathrm d\, \mathrm{log}_{10}N_{\upmu}^{\mathrm{tru}} \; \; \; ,
\end{split}
\end{equation}
\begin{equation}\label{eq:kernel_function}
\begin{split}
\mathrm{with} \; \; \;  
s_n&=s_n\left(  \mathrm{log}_{10}N_{\mathrm{ch}}^{\mathrm{tru}} ,  \mathrm{log}_{10}N_{\upmu}^{\mathrm{tru}}  \ | \  \mathrm{log}_{10}E           \right) \; \; \; ,\\
\varepsilon_n&=\varepsilon_n \left(  \mathrm{log}_{10}N_{\mathrm{ch}}^{\mathrm{tru}} ,  \mathrm{log}_{10}N_{\upmu}^{\mathrm{tru}}                              \right) \; \; \; ,\\
r_n&=r_n  \left(    \mathrm{log}_{10}N_{\mathrm{ch}}^{\mathrm{rec}}, \mathrm{log}_{10}N_{\upmu}^{\mathrm{rec}}                  \ | \   \mathrm{log}_{10}N_{\mathrm{ch}}^{\mathrm{tru}} ,  \mathrm{log}_{10}N_{\upmu}^{\mathrm{tru}}      \right) \; \; \; .
\end{split}
\end{equation}
More precisely, $s_n$ is the probability that a nucleus $n$, having an energy $E$, induces an air shower containing a specific number of charged particles $\mathrm{log}_{10}N_{\mathrm{ch}}^{\mathrm{tru}}$ and muons $\mathrm{log}_{10}N_{\upmu}^{\mathrm{tru}}$ when arriving at the detection plane. The probability to reconstruct, due to the resolution and possible systematic reconstruction uncertainties, a certain number of charged particles $\mathrm{log}_{10}N_{\mathrm{ch}}^{\mathrm{rec}}$ and muons $\mathrm{log}_{10}N_{\upmu}^{\mathrm{rec}}$, instead of the true ones $\mathrm{log}_{10}N_{\mathrm{ch}}^{\mathrm{tru}}$ and $\mathrm{log}_{10}N_{\upmu}^{\mathrm{tru}}$, is described by $r_n$. 

Equation~(\ref{eq:faltung_des_primaerspektrums}) can mathematically be understood as a system of coupled integral equations and is classified as a Fredholm integral equation of the first kind. In a straightforward way, this equation can be reformulated in terms of a matrix equation (cf.~\cite{lit:phd_fuhrmann} for the comprehensive calculation):
\begin{equation}\label{eq:faltung_matrix_equation}
\overrightarrow{Y}=\bm{R} \overrightarrow{X}\; \; \;,
\end{equation}
with the data vector $\overrightarrow{Y}$, whose elements $y_i$ are the cell contents $N_i$ in Eq.(\ref{eq:faltung_des_primaerspektrums}). The elements $x_i$ of the vector $\overrightarrow{X}$ represent the values of the sought-after differential energy spectra $\mathrm d J_n/\mathrm d\, \mathrm{log}_{10}E$, for all considered primaries $n$ consecutively. The conditional probabilities $p_n$ are included in the so-called transfer or response matrix $\bm{R}$, which relates the primary energy spectra to the measured shower sizes. 

There are various methods to solve such an equation, albeit resolvability often does not \textit{per se} imply uniqueness. It was found that the unfolding algorithm of Gold \cite{lit:gold_alg} yields appropriate and robust solutions. It is an iterative procedure and \textit{de facto} related to a minimization of a chi-square function. For countercheck purposes, all results are validated in \cite{lit:phd_fuhrmann} by means of two additional algorithms: one is an iterative method applying Bayes' theorem \cite{lit:bayes_alg}, which also performs high stability, and the other is a regularized unfolding based on a combination of the least-squares method with the principle of reduced cross-entropy \cite{lit:entropie_alg}, which yields slightly poorer results. For more details about the algorithms used, cf. \cite{lit:phd_fuhrmann,lit:kascade-unfolding}. 

All these solution strategies have in common that the response matrix $\bm{R}$, and therefore the response function $p_n$, have to be known \textit{a priori}. 
\section{Determination of the response matrix}
\label{Sec:reponse_matrix}
\begin{figure*}[p!]
 \centering
 \includegraphics[width=0.9\columnwidth,bb= 0 0 567 459]{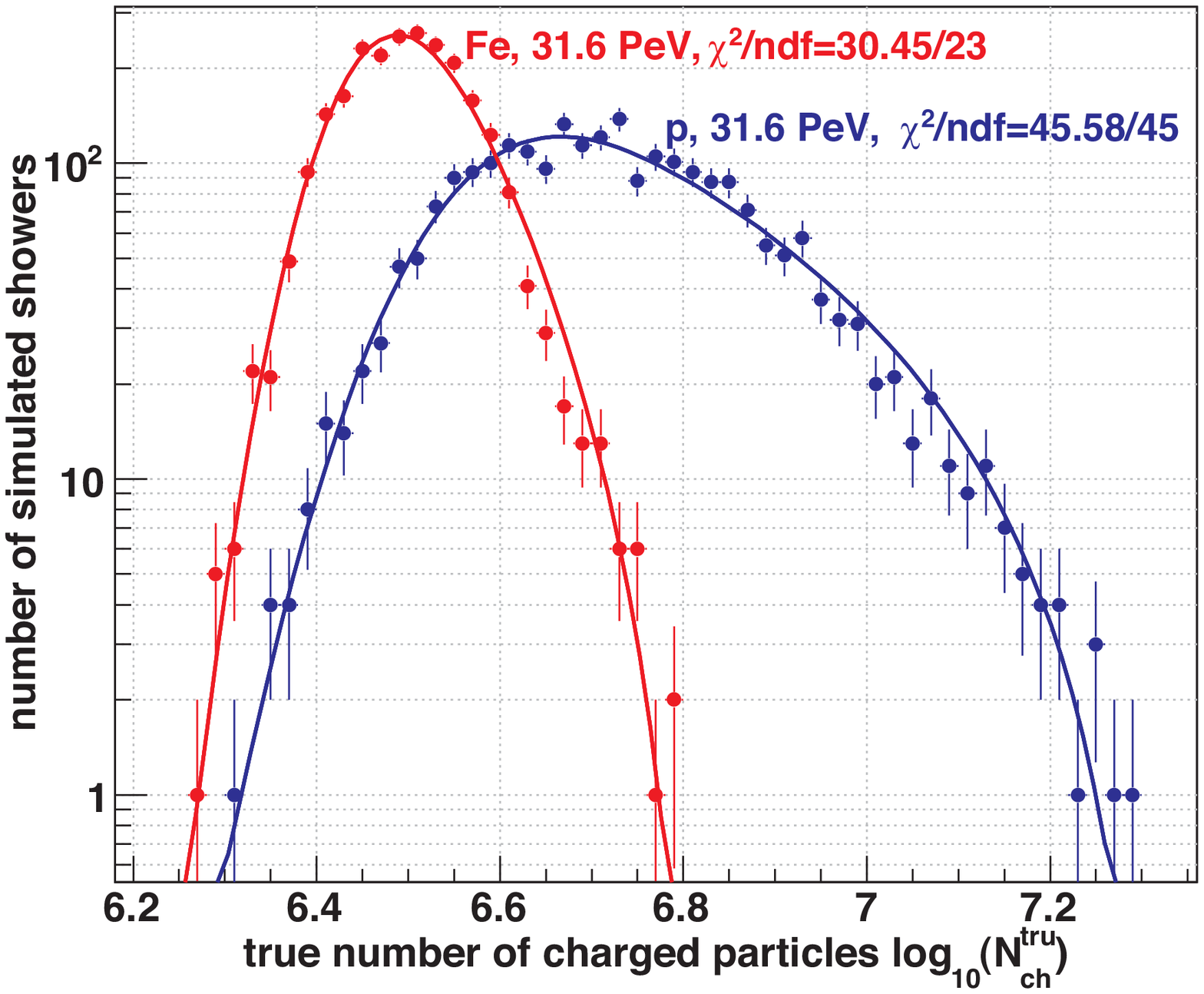}
 \includegraphics[width=0.9\columnwidth,bb= 0 0 567 459]{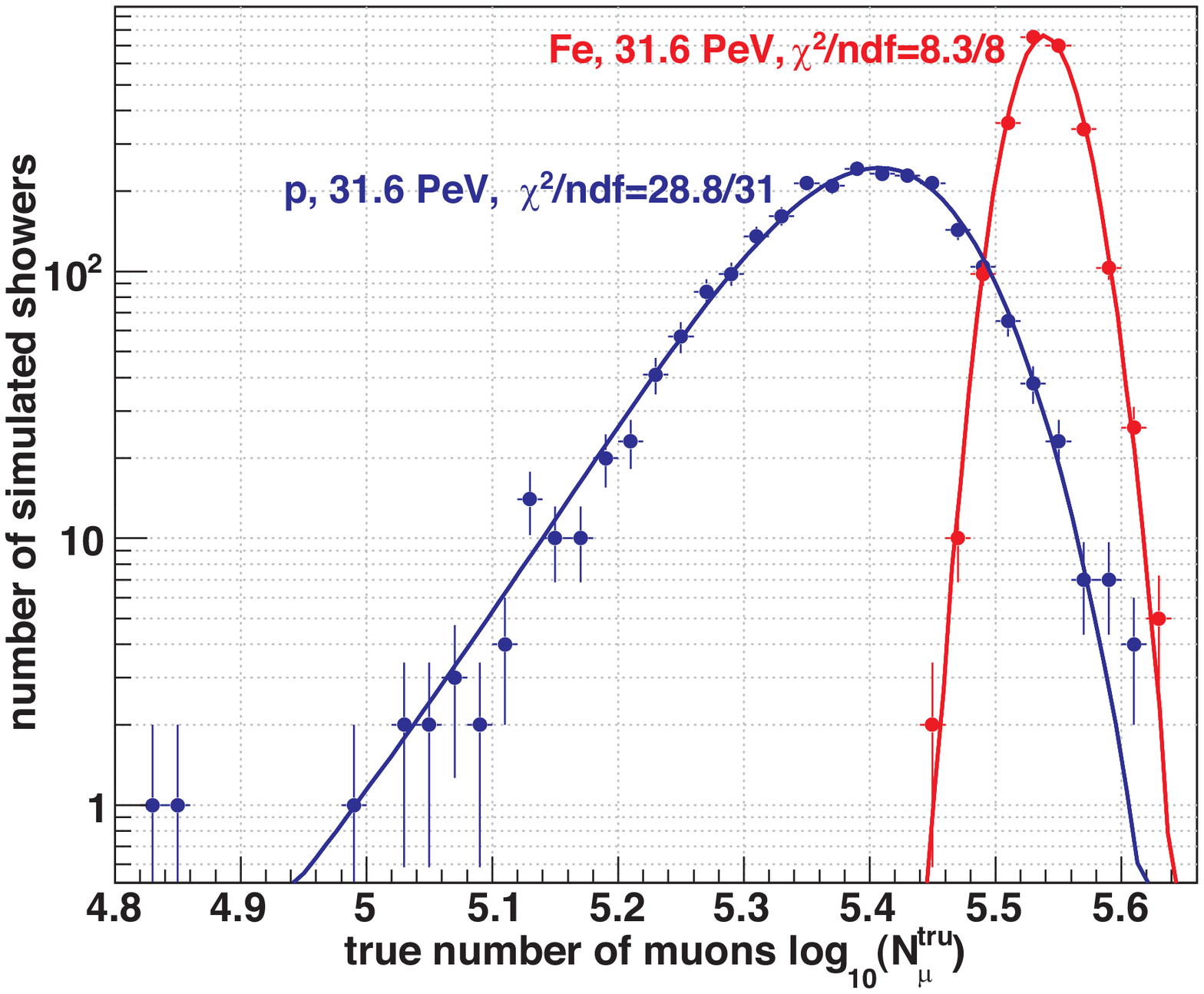}
 \caption{The distribution of the charged particle number (left) and muon number (right) at observation level based on the simulations, exemplarily shown for proton (blue) or iron (red) initiated showers with a primary energy of $31.6$~PeV. The distributions are parametrized using appropriate functions (curves), whereby in case of the muon distribution the fit function used takes account of the correlation between the two shower sizes.}
 \label{fig:showerfluc}
\end{figure*}

The calculation of the matrix elements of the response matrix $\bm{R}$, i.e. the determination of the quantities $s_n$, $\varepsilon_n$, and $r_n$ of Eq.(\ref{eq:pn}), is realized with Monte Carlo simulations. The simulated distributions are parametrized in order to simplify the mathematical integrations, as well as to apply some kind of smoothing necessary due to the limited Monte Carlo statistics. Furthermore, a conditioning is applied to the response matrix. Even though the considered number of primary cosmic ray mass groups is restricted to only five (represented by protons (p), as well as helium (He), carbon (C), silicon (Si), and iron (Fe) nuclei), their simulated distributions for the intrinsic shower fluctuations already overlap to a large extent (cf. Fig.~\ref{fig:showerfluc}, where even the distributions of protons and of iron nuclei are overlapping). This is again worsened by the additional smearing due to the reconstruction resolution. Therefore, the response matrix is almost singular, and hence Eq.(\ref{eq:faltung_matrix_equation}) states an ill-conditioned problem so that a straightforward solution by a matrix inversion would yield meaningless results. The applied unfolding algorithms, however, allow reliable solutions under the premise that the matrix equation exhibits a minimum level of stability, i.e. given the case that the response matrix is sufficiently conditioned. The so-called condition number of a response matrix is given by the ratio of the largest to the smallest singular value of the matrix\footnote{More precisely, since the response matrix is not invertible in this case, a singular value decomposition is performed to compute the pseudoinverse, and, finally, the condition number (see~\cite{lit:phd_fuhrmann}).}. To ensure the statistical significance of the solution, the condition number should not exceed\footnote{The optimal maximum value was determined using test spectra and trying different values.} $10^{7}$ in our case, requiring that no more than five primary nuclei are taken into account, and that only probabilities $p_{n}$ larger than $10^{-4}$ contribute to the response matrix. On the other hand, investigations based on a Kolmogorov-Smirnov and a chi-square test (analogous to the tests presented in Section~\ref{Sec:discussion_quality}) have shown that at least five primary mass groups are needed in order to describe the measured data sufficiently. Hence, five primary nuclei will be considered in this work (see~\cite{lit:phd_fuhrmann} for further details). 

For the parametrization of the intrinsic shower fluctuations, the development of air showers is simulated by means of CORSIKA~\cite{lit:HeckCorsika}, version 6.307, based on the interaction models \mbox{QGSJET-II-02}~\citep{lit:qgsjet-ii-model-zitat1, lit:qgsjet-ii-model-zitat2} and \mbox{FLUKA} 2002.4~\citep{lit:FlukaAllgemeinZitat1,lit:Fluka2002_4_Zitat1,lit:FlukaAllgemeinZitat2}. For each of the five primaries separately, the two-dimensional shower size distribution is simulated and parametrized for thinned\footnote{In order to get sufficient simulation statistics in a certain amount of time, the thinning option~\citep{lit:HeckCorsikaThinningEtc} of CORSIKA was enabled in order to save computing time. The selected thinning level $10^{-6}$ means that for interactions, where particles with energies less than $10^{-6}$ of the primary particle energy are generated, only one particle is kept and is assigned a weight that accounts for the energy of the neglected particles. It was found that this will not have any significant impact on the shower size distributions used, and hence on the analysis, as detailed investigations in \cite{lit:kascade-unfolding} have proved.} air showers with cores distributed uniformly over an area slightly larger than the KASCADE-Grande detector field, and with isotropically distributed zenith angles $\leq 18^\circ$. The CORSIKA simulations are mono-energetic in order to get sufficiently large statistics for the parametrizations, and to avoid an \textit{a priori} assumption of a specific index of the power law spectrum of cosmic rays. In order to enhance the quality, instead of fitting the correlated two-dimensional $\mathrm{log}_{10}N_{\mathrm{ch}}^{\mathrm{tru}}$-$\mathrm{log}_{10}N_{\upmu}^{\mathrm{tru}}$-distribution immediately, it is done in two steps. Firstly, the distribution of charged particles is parametrized (cf. example in Fig.~\ref{fig:showerfluc}, left panel) using an appropriate one-dimensional fit function, thereafter the one of muons (cf. example in Fig.~\ref{fig:showerfluc}, right panel).
\begin{figure*}[p!]
 \centering
 \includegraphics[width=0.9\columnwidth,bb= 0 0 567 459]{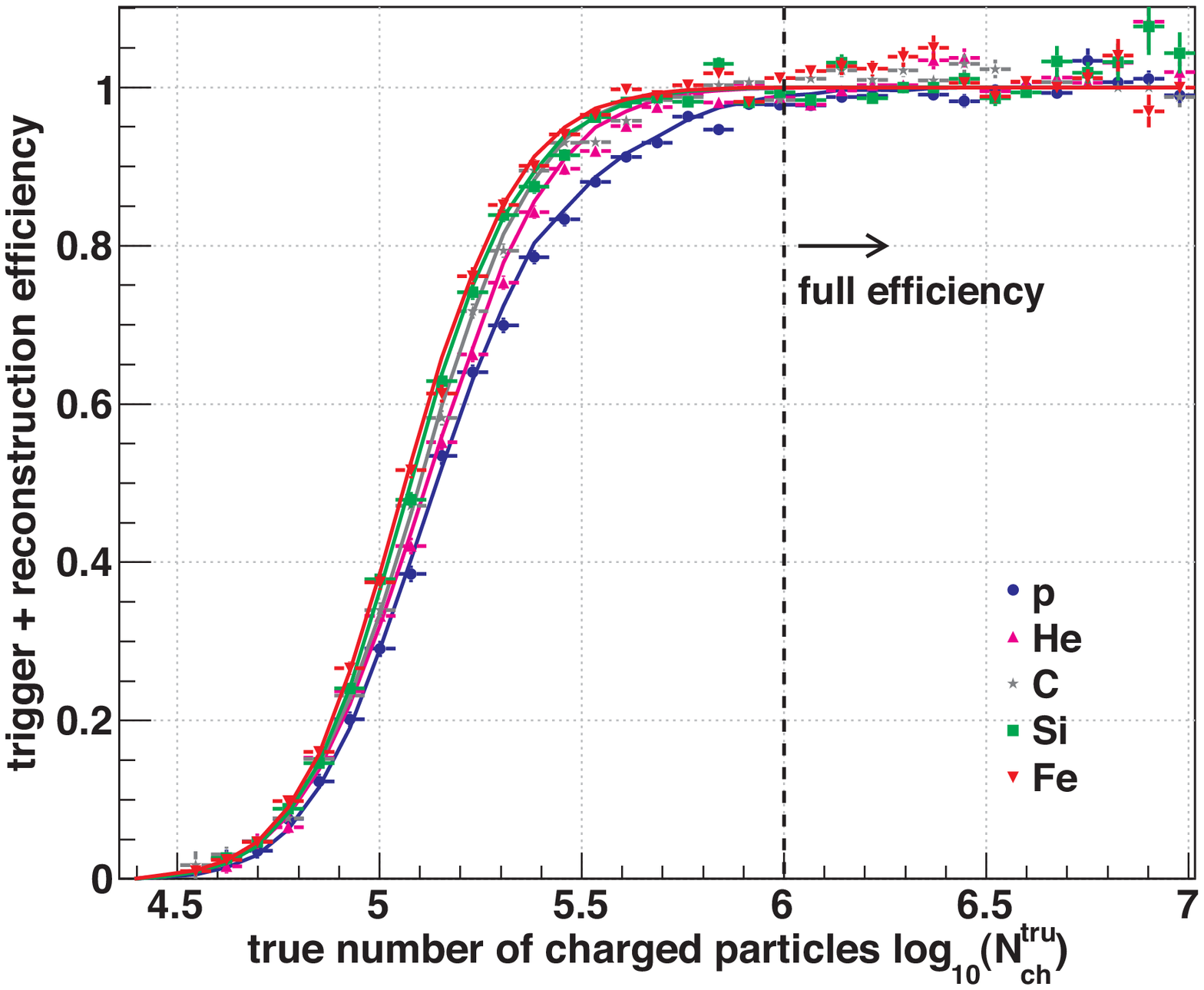}
\includegraphics[width=0.9\columnwidth,bb= 0 0 567 459]{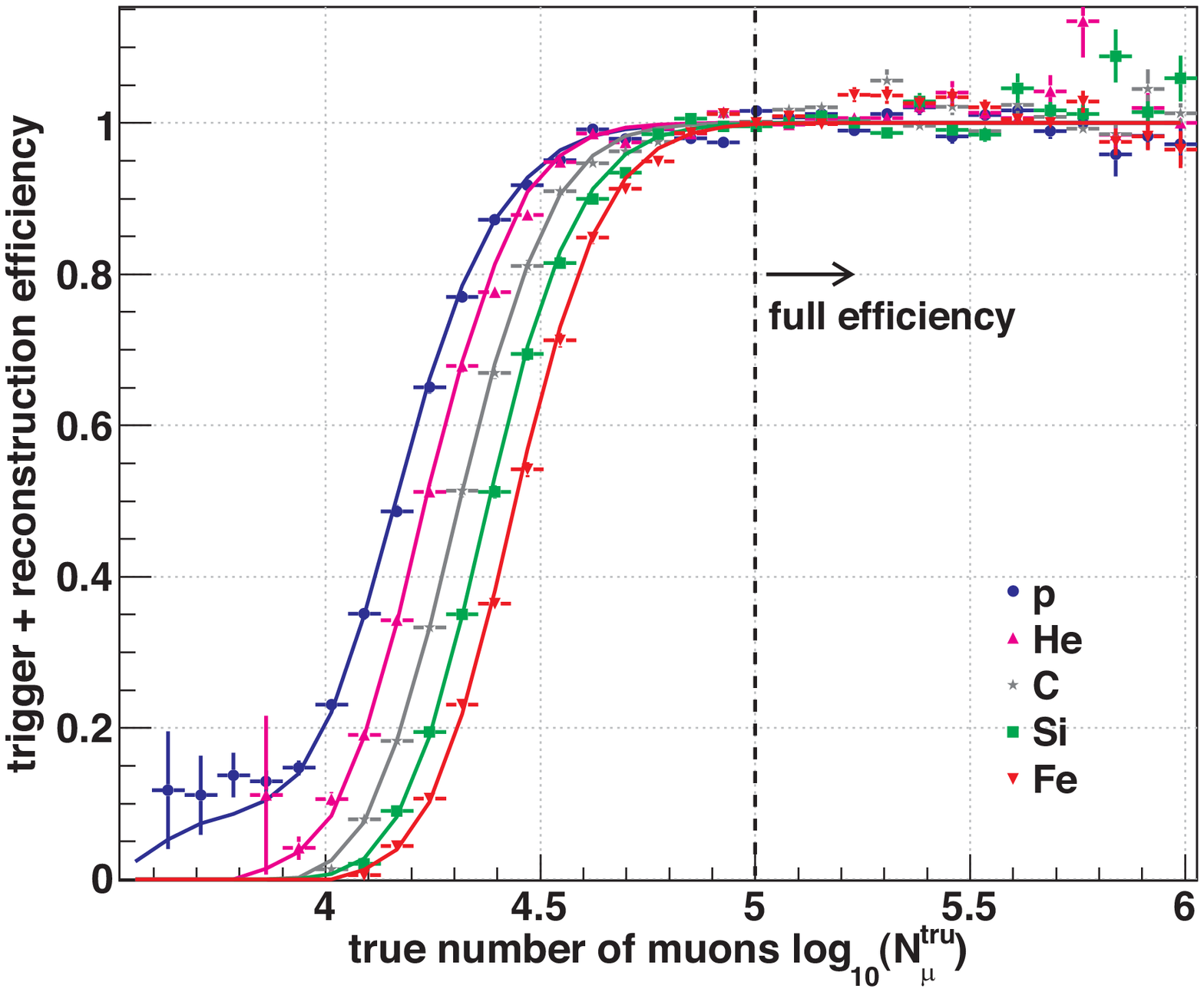}
 \caption{The efficiency computed based on the simulations in dependence on the true number of charged particles (left) and muons (right) for all five considered primaries, as well as the determined parametrizations (curves).}
 \label{fig:efficiency}
\end{figure*}
\begin{figure*}[p!]
 \centering
 \includegraphics[width=0.9\columnwidth,bb= 0 0 567 459]{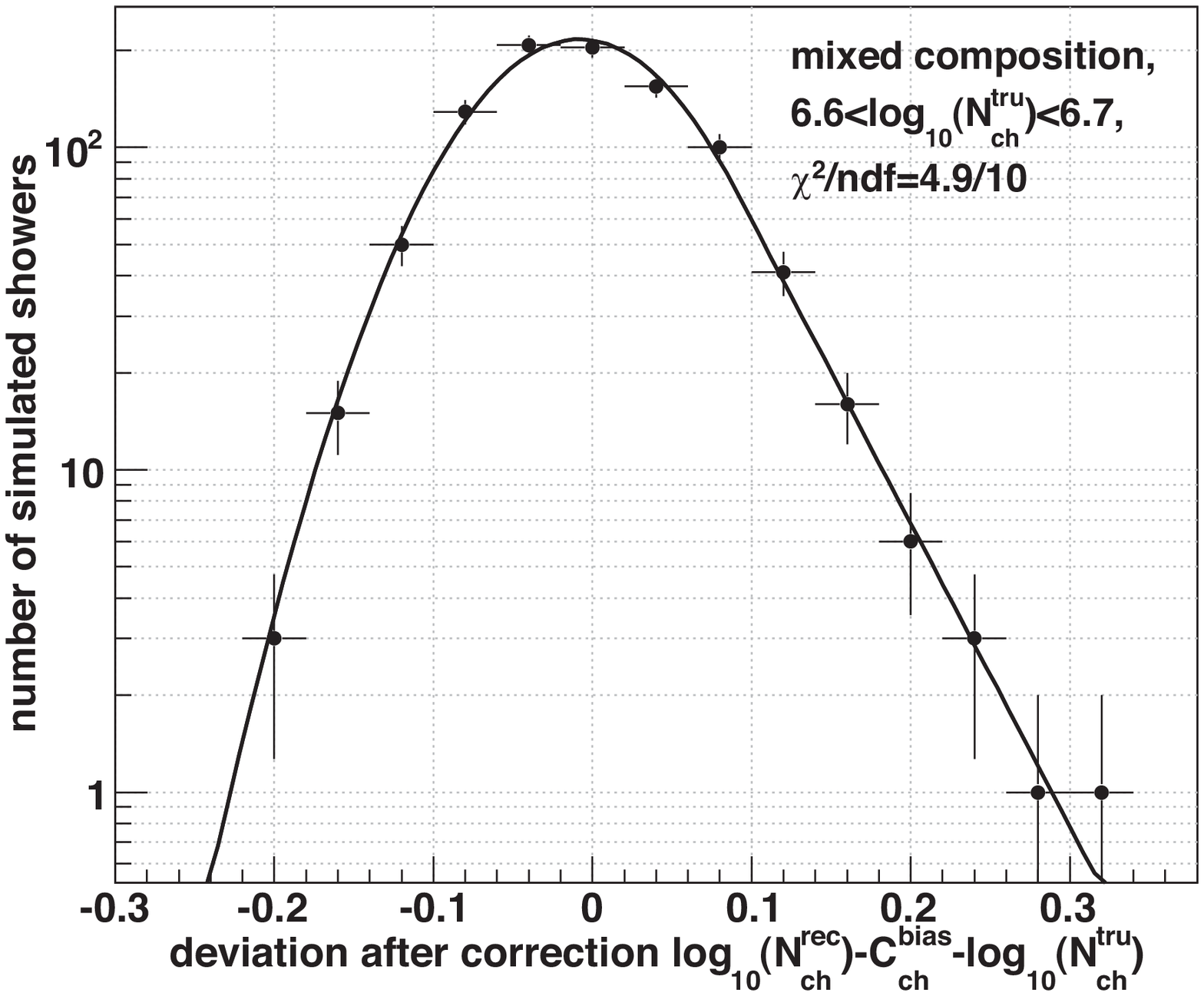}
  \includegraphics[width=0.9\columnwidth,bb= 0 0 567 459]{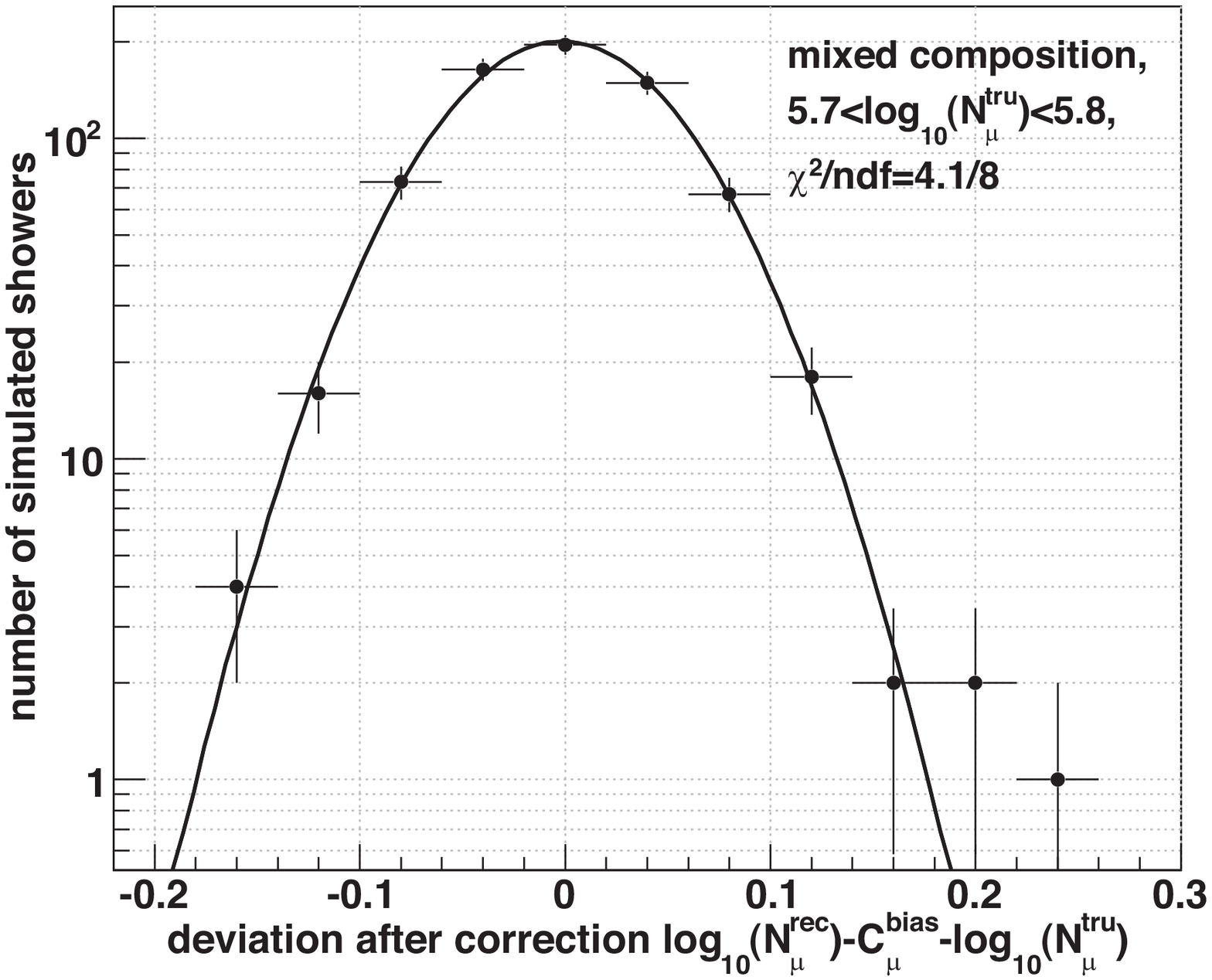}
 \caption{The distribution of the deviations between the reconstructed (bias corrected) and true shower sizes in case of charged particles (left, exemplarily for the interval  $6.6<\log_{10}(N_{\mathrm{ch}}^{\mathrm{tru}})<6.7$) and of muons (right, exemplarily for the interval $5.7<\log_{10}(N_{\upmu}^{\mathrm{tru}})<5.8$), as well as the determined parametrization (curves). To increase the available simulation statistics, a mixed composition is used.}
 \label{fig:recaccuracy}
\end{figure*}
In the latter case, the two-dimensional fit function describing the $\mathrm{log}_{10}N_{\mathrm{ch}}^{\mathrm{tru}}$-$\mathrm{log}_{10}N_{\upmu}^{\mathrm{tru}}$-distribution is used. Since the parameters of the parametrization of the distribution of the charged particles are known from step one, the two-dimensional fit function can be transferred to a one-dimensional one by means of integration over the charged particle number, such that the remaining parameters describing the muon distribution part can be determined by fitting the one-dimensional muon number distributions. Thereby, the correlation between the number of charged particles and that of muons is considered in the fit procedure (this is explained in more detail in \cite{lit:phd_fuhrmann}). The parameters of the parametrizations determined at the discrete energies are finally interpolated in order to extrapolate the parametrization to a continuum. The simulated energies are: 2~PeV, 5~PeV, 10~PeV, $31.6$~PeV, 100~PeV, 316~PeV, 1000~PeV, and 3160~PeV; the numbers of simulated showers are: 6400, 4800, 3200, 2400, 1600, 1200, 800, and 400, respectively. 

The efficiency as well as the resolution and systematic uncertainties for the five primaries are simulated using CRES\footnote{\underline{C}osmic \underline{R}ay \underline{E}vent \underline{S}imulation, a program package developed for the KASCADE~\cite{lit:kascade_allgemein_nimpaper} detector simulation.}, which bases on the GEANT~3.21~\cite{lit:GEANT,lit:GEANT3_21} detector description and simulation tool. A second set of unthinned\footnote{CRES handles only unthinned showers.} air showers simulated with CORSIKA (again based on \mbox{QGSJET-II-02} and \mbox{FLUKA} 2002.4) serves as input for CRES. Unlike in case of the intrinsic shower fluctuations where mono-energetic simulations are used, now a continuous energy spectrum following a power law with differential index $-2$ and comprising energies from 0.1~PeV to 3160~PeV is assumed. This spectrum is roughly one order of magnitude harder than the one actually measured, but is representing a compromise between sufficient statistics at the highest energies and computing time. Later, the simulated spectrum is reweighted to one with index $-3$. It was found that the obtained parametrizations do not differ significantly if, alternatively, indices of $-2.7$ or $-3.3$ are assumed, so that the exact value of the index is of minor importance. With about $353\, 000$ simulated events per primary, the statistic of the simulations is roughly comparable to the one given in the measured data sample before applying quality cuts.

The combined trigger and reconstruction efficiency, simply called ``efficiency'', and its parametrization is depicted in Fig.~\ref{fig:efficiency}. Full efficiency of the experiment is given for $\log_{10}(N_{\mathrm{ch}}^{\mathrm{tru}})\geq6.0$ and $\log_{10}(N_{\upmu}^{\mathrm{tru}})\geq5.0$. Since in this analysis only measured air showers with shower sizes beyond the threshold of full efficiency are considered, the goodness of the parametrization of the efficiency is of minor importance\footnote{However, for the computation of the response matrix the parametrization of the efficiency is necessary, since shower sizes below the threshold of full efficiency are regarded to account for possible migration effects caused by the intrinsic shower fluctuations.}. 

In order to parametrize the dependence of the resolution of the experiment on the true sizes, a possible bias in the charged particle and muon number reconstruction must first be corrected by using appropriate correction functions $C_{\mathrm{ch}}^{\mathrm{bias}}$ and $C_{\upmu}^{\mathrm{bias}}$, respectively, determined based on the simulations. The correction is typically in the order of less than $10\%$. The distributions of the remaining deviations between the reconstructed (and bias corrected) and true shower sizes are depicted in Fig.~\ref{fig:recaccuracy} for the charged particle number (left panel) and for the muon number (right panel), in case of discrete exemplary true shower size intervals (corresponding to about 30~PeV to 40~PeV primary energy). Since the resolution does not differ significantly between different primaries, in order to increase statistics, the simulations for the five primary particles can be combined to a mixed composition set serving for the parametrization. 

\begin{figure}
 \centering
 \includegraphics[width=0.985\columnwidth,bb= 0 0 567 459]{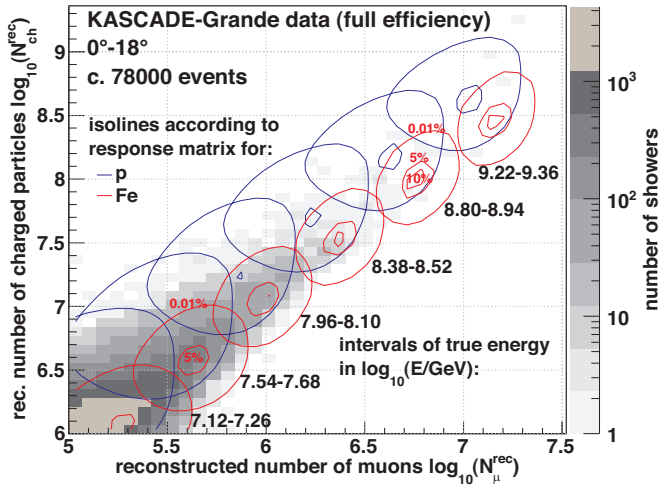}
 \caption{A comparison between the measured shower size distribution (grey histogram) and some isolines with $\log_{10}(N_{\upmu}^{\mathrm{rec}})$-$\log_{10}(N_{\mathrm{ch}}^{\mathrm{rec}})$ combinations of constant probability according to the parametrizations (from the inner to the outermost line 10\%, 5\% and 0.01\%). This is illustrated exemplarily for protons as well as iron nuclei, and in case of six energy bins (labelled below each isoline set).
}
 \label{fig:response_vs_real_shower_size_contour}
\end{figure}

In Fig.~\ref{fig:response_vs_real_shower_size_contour},
the measured shower size plane is compared to the probabilities given by the final response matrix taking into account the entire parametrizations, i.e. that of the intrinsic shower fluctuations as well as that of the properties of the experiment. Shown are some isolines representing the cells $\left( \log_{10}(N_{\mathrm{ch}}^{\mathrm{rec}}), \log_{10}(N_{\upmu}^{\mathrm{rec}}) \right)_{i}$ of the data plane with constant probability (from the inner\footnote{In case of smaller energies, the widths of the probability distributions are as large that there are no individual probabilities larger than 0.1 or even 0.05, such that the inner isolines are missing in these cases.} to the outermost isoline: $0.1$, $0.05$ and $10^{-4}$ probability density). For reasons of clarity, only the results for two exemplary primaries are illustrated: protons and iron nuclei. The isolines, which correspond to the $\log_{10}(N_{\mathrm{ch}}^{\mathrm{rec}})$-$\log_{10}(N_{\upmu}^{\mathrm{rec}})$ combinations with a probability of $10^{-4}$, represent the smallest probability value just considered in the response matrix after its conditioning. As can be seen, these outer isolines cover almost all measured data; hence, the minimal probability is not set too large. 
\section{Error propagation}
\label{Sec:error_propagation}
The determination of the elemental energy spectra will be subjected to influences of different error sources. They can roughly be classified in four categories (cf.~\cite{lit:phd_fuhrmann} for details): 
\begin{enumerate}[(i)]
 \item {
\textit{Statistical uncertainties due to the limited measurement time:}
Due to the limited exposure, the measured data sample will suffer from unpreventable statistical uncertainties, which are expected to be Poisson distributed. These uncertainties will be propagated through the applied unfolding algorithm and are usually amplified thereby. The statistical uncertainties can be determined by means of a frequentist approach: The measured two-dimensional shower size plane is considered as probability distribution. Based on a random generator, a couple of artificial data sets are generated, which are unfolded individually. The spread of the solutions represents a good estimate for the statistical uncertainty due to the limited measurement time.
}
\item {
\textit{Systematic bias induced by the unfolding method:}
In the context of the convergence properties of the iterative unfolding algorithms, small numbers of iteration steps will on the one hand reduce the amplification of the statistical uncertainties of the data sample, and on the other hand will result in a solution that is deviating from the exact one. In case of the regularized techniques it is similar, since the regularization damps oscillations, but, conversely, results in a biased solution. In this work, the number of iteration steps, respectively the regularization parameter, is chosen such that an optimal balance between the statistical uncertainties and the systematic bias is achieved. The bias can be estimated based on the principle of the bootstrap methods: The measured two-dimensional shower size plane is unfolded under a certain number of iteration steps. Based on the derived solution and using a random generator, while the response matrix contributes the respective probability distribution, a couple of toy data sets can be generated. Unfolding them and comparing the solutions to the original solution yields an estimate for the mean bias induced by the unfolding algorithm for this specific number of iteration steps. 
}
\item {
\textit{Systematic uncertainties due to the limited Monte Carlo statistics:}
Due to limited computing time, only Monte Carlo simulation sets with limited statistics can be generated resulting in an uncertainty in the determination of the response matrix. Furthermore, the conditioning that was applied to the response matrix has systematic impacts, which are, however, small and can be neglected in this work. The systematic uncertainties of the response matrix will finally affect the unfolded solution. The influence of the limited Monte Carlo statistics can be examined by generating further sets of response matrices used for unfolding the measured data set. First, the parameters of the parametrizations can be varied within their statistical precision. However, the effect was comparatively small, meaning that the simulation statistics are basically large enough. Second, the tails of the parametrizations can be varied within the statistical accuracy of the simulated distributions. While from a pure statistical point of view the quality of the fits does not change, from a physical perspective the exact knowledge of the tails is of high importance in order to account for the bin-to-bin migration effects in combination with the steeply falling spectrum of cosmic rays. By varying the parametrizations as extensively as possible given the statistical accuracy, at least a maximal range of systematic uncertainty caused by the limited Monte Carlo statistics can be estimated. 
}
\item {
\textit{Systematic uncertainties due to the systematic uncertainty in the Monte Carlo simulations:}
The Monte Carlo simulations used to compute the response matrix are based on the high energy interaction model \mbox{QGSJET-II-02}~\citep{lit:qgsjet-ii-model-zitat1, lit:qgsjet-ii-model-zitat2} and the low energy interaction model \mbox{FLUKA} 2002.4~\citep{lit:FlukaAllgemeinZitat1,lit:Fluka2002_4_Zitat1,lit:FlukaAllgemeinZitat2}. \citet{lit:alle_high_energy_modelle_haben_probleme} compared the first Large Hadron Collider (LHC) data with the predictions of various Monte Carlo event generators, including e.g. the models \mbox{QGSJET 01}~\cite{lit:qgsjet01}, \mbox{QGSJET-II}~\cite{lit:qgsjet-ii-model-zitat1}, \mbox{SIBYLL}~2.1~\cite{lit:sibyll}, and \mbox{EPOS}~1.99~\cite{lit:epos}. They concluded that there is basically a reasonable overall agreement; but, they stated also that none of the investigated models can describe consistently all measured observables at the LHC. Nevertheless, it was found that the model QGSJET-II-02 yields results, which agree with the data measured with KASCADE-Grande; and hence it can be expected that the result is not far off the truth (cf. Section~\ref{Sec:discussion_quality}). A possible deficient description of the contributing physical processes would result in systematic errors in the response matrix, finally leading to a wrong result of the deconvolution. These uncertainties are difficult to quantify as all models can fail if new physics is appearing in this energy range. However, in \cite{lit:kascade-unfolding,lit:KG_all_part_spec_paper} it was shown that the high energy interaction model affects primarily the relative abundances of the mass groups and the absolute scale in energy assignment, while specific structures in the spectra are conserved. For example, with \mbox{EPOS}~1.99 the energy assignment for an individual
event is by approximately 10\% lower than interpreted with \mbox{QGSJET-II-02}, while for \mbox{SIBYLL 2.1} the energy is 10\% higher \cite{lit:KG_all_part_spec_paper,lit:KG_influence_hadr_int}. In addition, it is known that the low energy interaction model has less influence on the final result, as already the analyses based on the KASCADE measurements have proved \cite{lit:KASCADE_low_energy_interaction_model}.  
}
\end{enumerate}
\section{Monte Carlo tests}
\label{Sec:monte_carlo_tests}
Before the unfolding techniques are applied to the measured data, the whole procedure is tested with simulations. For that purpose, different models for the energy spectra for the elemental groups of cosmic rays can be assumed. Based on these models and using a random generator, a sample of elemental test energy spectra can be generated. By means of the random generator, statistical fluctuations are introduced equivalent to ones a measurement suffers from due to the limited measurement time. Applying a random generator once more, and, this time, using the entries of the response matrix as probability distributions, a toy 
\begin{figure*}[t]
 \centering
 \includegraphics[width=0.9\columnwidth,bb= 0 0 567 484]{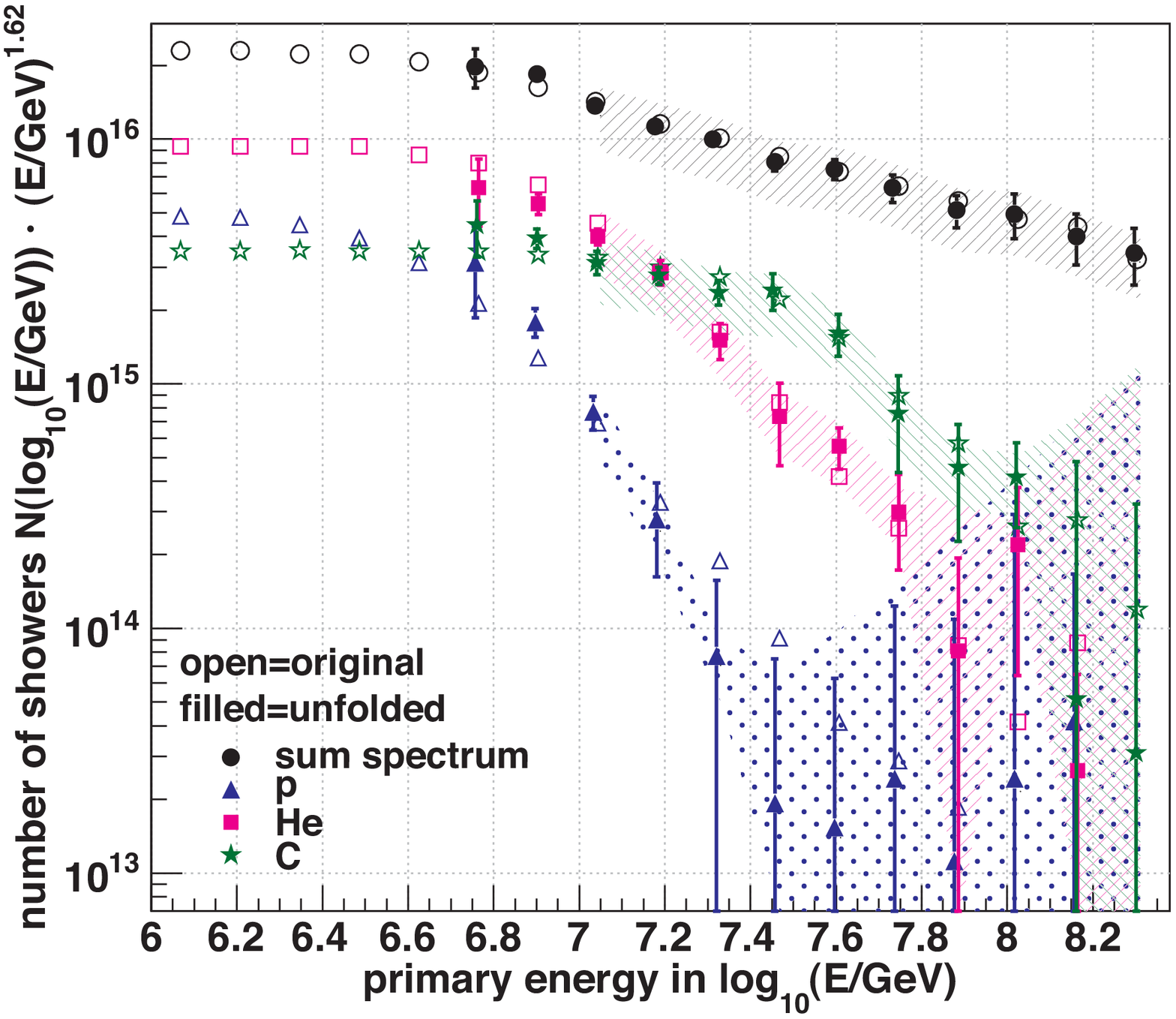}
 \includegraphics[width=0.9\columnwidth,bb= 0 0 567 484]{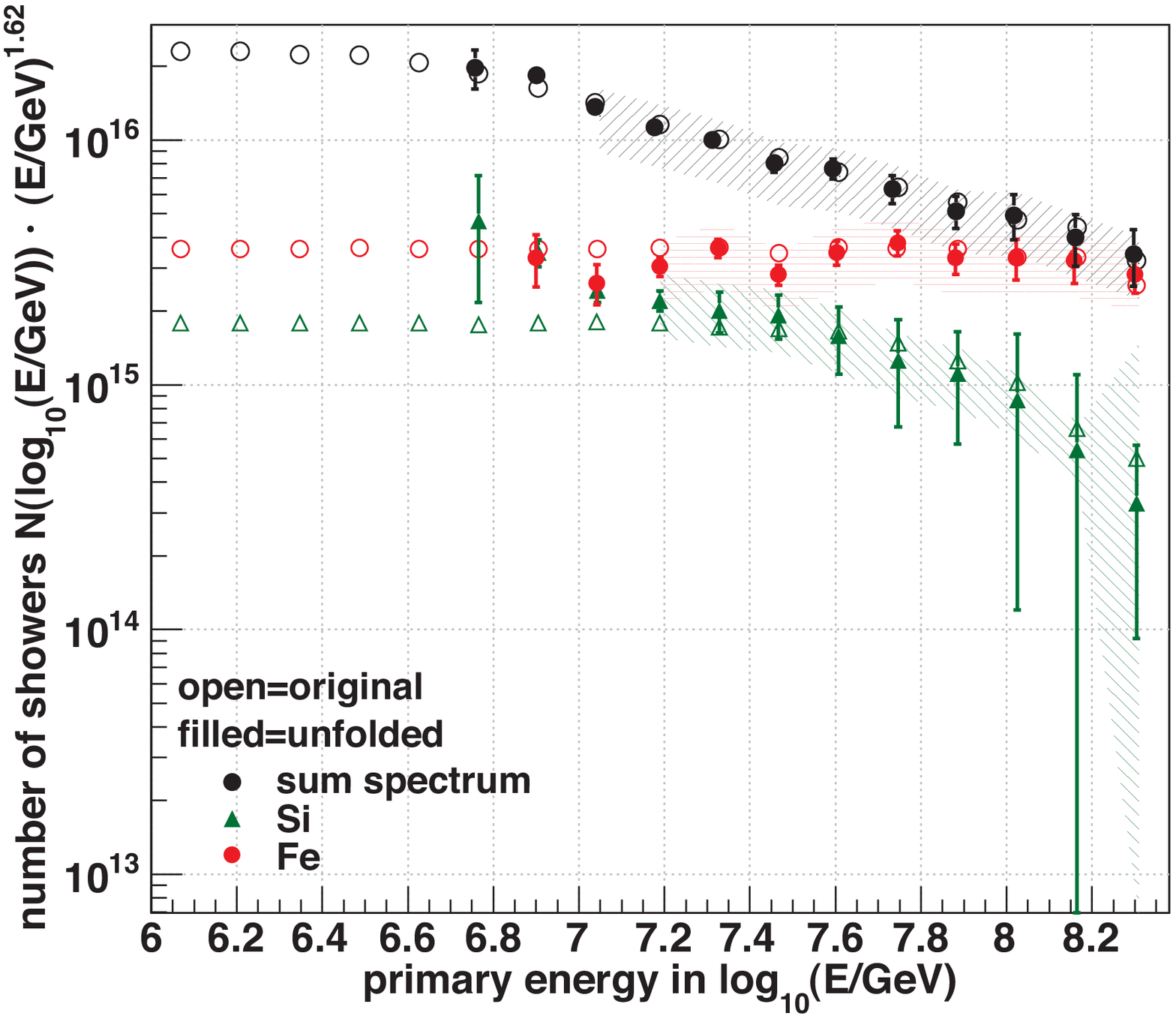}
 \caption{Realistic test energy spectra (open symbols) are shown together with the spectra determined by an unfolding (based on Gold's unfolding algorithm \cite{lit:gold_alg}) of the toy data set generated based on these test spectra (filled symbols): on the left panel, in case of the lighter mass groups, and, on the right panel, for the heavier ones. The sum spectrum, which is the sum of all five unfolded elemental spectra, is also depicted. The error bars are the statistical uncertainties due to the limited measurement time, the error bands represent the systematic bias induced by Gold's unfolding algorithm. The bands are only shown for energy ranges, where the estimation of the systematic uncertainties works reliably.}
 \label{fig:testspek}
\end{figure*}
data set (i.e. a two-dimensional shower size spectrum) can be generated from the elemental test energy spectra. The assumed exposure is typically chosen such that the number of entries in the generated two-dimensional toy shower size spectrum is comparable to that in case of the measurement. The artificial data sets are used to test the unfolding procedure. 

We have performed such tests for a large number of toy energy spectra, generated under different model assumptions including energy spectra with and without knee-like structures and varying relative abundances of the primary mass groups. For example, test energy spectra with primaries of equal abundance were used in order to check whether or not the applied unfolding technique favours a specific cosmic ray mass group. Another approach was to consider test energy spectra following a single power law, in order to rule out that possible knee-like structures in the unfolded spectra are caused by the unfolding method itself. In all these test scenarios, equally good results for the unfolding technique were achieved. Within systematic uncertainties, the unfolding results were always compatible with the input spectra. Similar tests were performed by unfolding spectra built up by different number of primary mass groups. The tests resulted in the conclusion that the resolution of our measurements and the fluctuations in the data allow to unfold in five mass groups. Details of such tests are described in~\cite{lit:kascade-unfolding,lit:phd_fuhrmann}. 

In the following, exemplarily, the results based on test spectra that are expected to be close to reality will be discussed. Their parameters are determined by fitting the energy spectra measured by the former KASCADE experiment \cite{lit:kascade-unfolding,lit:daniel_Bindig_diplom}. In Fig.~\ref{fig:testspek},
the generated test spectra are depicted (open symbols) in comparison to the ones obtained by an unfolding (filled symbols, based on Gold's unfolding algorithm \cite{lit:gold_alg}) of the toy shower size spectrum generated based on these test spectra. The error bars are the statistical uncertainties due to the limited measurement time propagated through the unfolding algorithm, while the error bands represent the systematic bias caused by Gold's unfolding algorithm. Since the response matrix is used for the generation of the toy data set as well as in the unfolding procedure itself, systematic uncertainties of the response matrix do not have any recognizable influence in these particular Monte Carlo tests, and are hence not included in the error bands. Basically, the unfolding method seems to yield good results, and reproduces specific structures in the spectra successfully. The artificial wobbling at low energies, especially in case of the spectrum of iron nuclei, can be explained as a systematic relic of the unfolding algorithm, and contributes to the systematic uncertainties. Furthermore, the procedure tends to overestimate the fluxes at higher energies due to the small number of true events, close to zero, in combination with the positive definiteness of Gold's unfolding algorithm. However, the energy ranges suffering from this problem are explicitly tagged unreliable by the estimated systematic and/or statistical uncertainties. In case of lower energies, larger systematic deviations between the original and the unfolded spectra for the heavier mass groups, represented by silicon and iron nuclei, can be observed. They are caused by the different convergence rates of Gold's algorithm below the threshold of full detection 
\begin{figure*}[t]
 \centering
 \includegraphics[width=0.9\columnwidth,bb= 0 0 567 484]{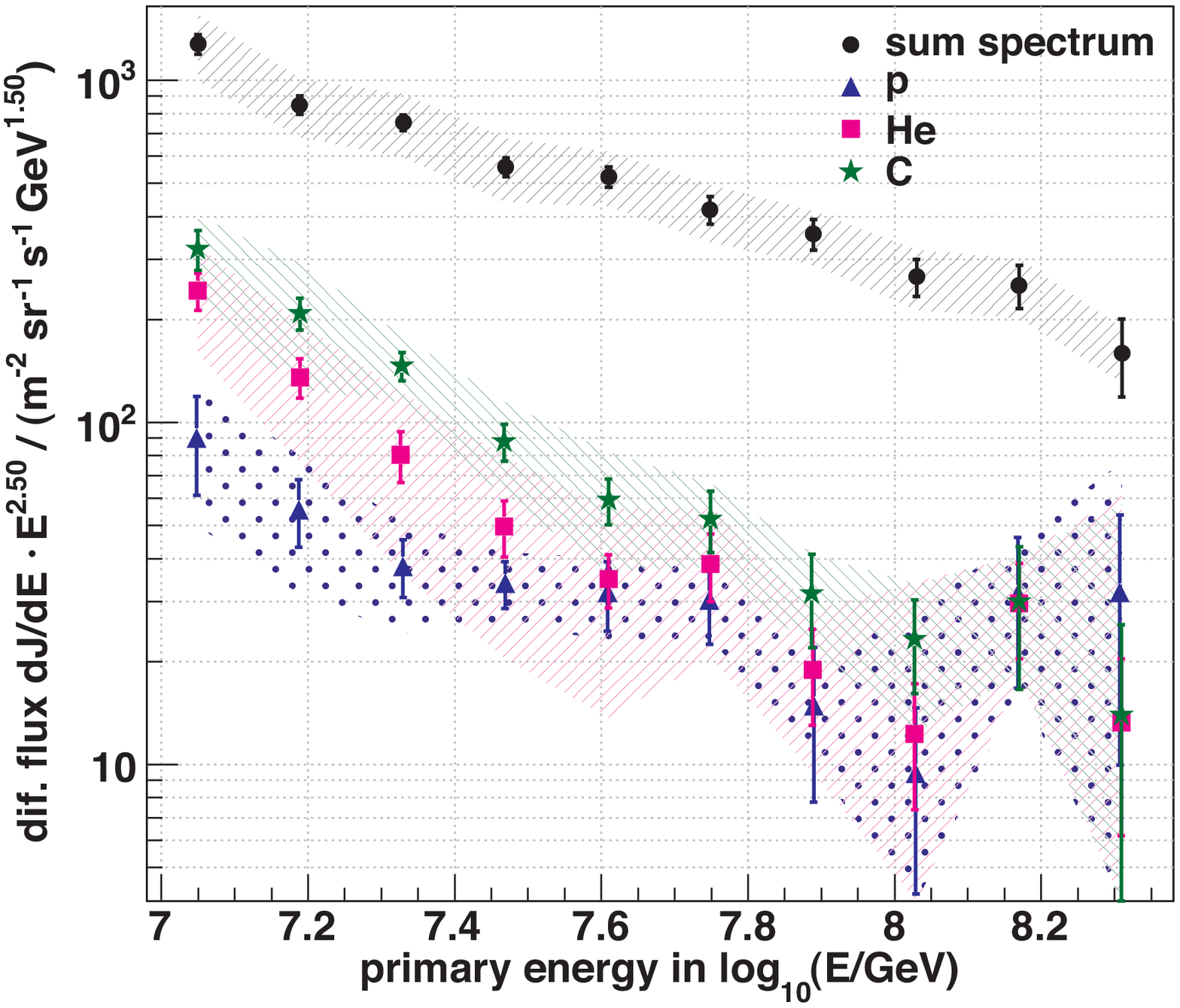}
 \includegraphics[width=0.9\columnwidth,bb= 0 0 567 484]{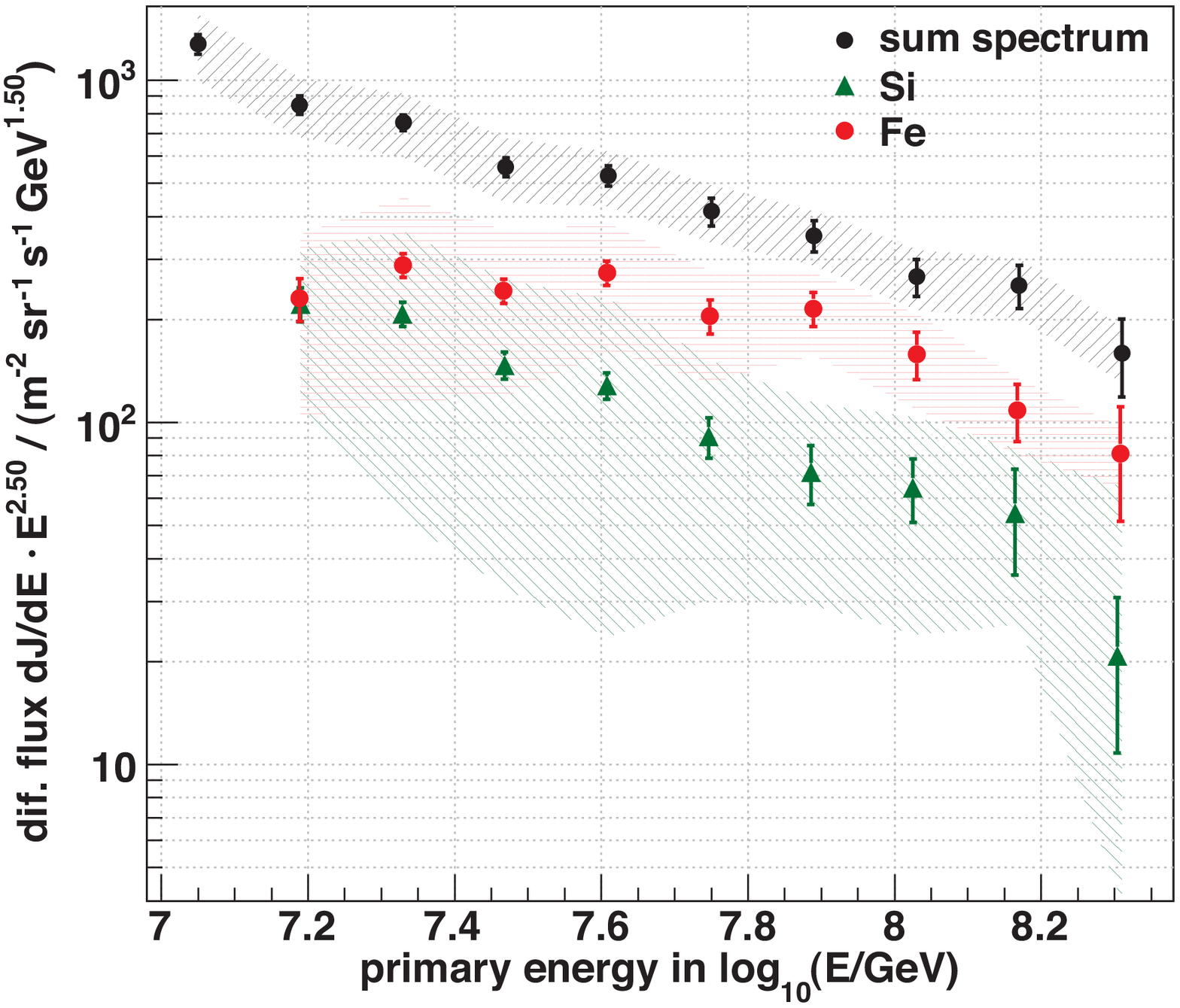}
 \caption{The unfolded energy spectra for elemental groups of cosmic rays, represented by protons, helium, and carbon nuclei (left panel) as well as by silicon and iron nuclei (right panel), based on KASCADE-Grande measurements. The all-particle spectrum that is the sum of all five individual spectra is also shown. The error bars represent the statistical uncertainties, while the error bands mark the maximal range of systematic uncertainties. The response matrix used bases on the interaction models \mbox{QGSJET-II-02}~\citep{lit:qgsjet-ii-model-zitat1, lit:qgsjet-ii-model-zitat2} and \mbox{FLUKA} 2002.4~\citep{lit:FlukaAllgemeinZitat1,lit:Fluka2002_4_Zitat1,lit:FlukaAllgemeinZitat2}.}
 \label{fig:final_elemtal_spectra}
\end{figure*}
efficiency\footnote{The detection efficiency incorporates the combined trigger and reconstruction efficiency (which bases on the true shower sizes) with the probability that an air shower contributes to the shower size plane that is serving as basis for our analysis (i.e. that it passes the cut $\log_{10}(N_{\mathrm{ch}}^{\mathrm{rec}})\geq6.0$ and $\log_{10}(N_{\upmu}^{\mathrm{rec}})\geq5.0$, which bases on the reconstructed shower sizes).}. These systematic effects can hardly be estimated in case of real data. Hence, the only possibility will be to demand a sufficiently large detection probability, which is realized if the energy is larger than $\log_{10}(E/\mathrm{GeV})\approx 7.0$ for the lighter mass groups, and larger than $\log_{10}(E/\mathrm{GeV})\approx 7.2$ for silicon and for iron. Hence, for energies below these thresholds, i.e. where the estimation of the systematic uncertainties is not comprehensive due to the missing uncertainty caused by the different convergence properties, no error bands are depicted in Fig.~\ref{fig:testspek}. In order to guarantee that the presented spectra are reliable within the given uncertainties, in the main analysis, energy ranges below these limits will be omitted in the depictions completely, though they are considered mathematically within the unfolding process itself. Considering all these insights, the unfolding procedure can be applied successfully to the measured shower size spectrum. 
\begin{figure}[b!]
 \centering
 \includegraphics[width=0.985\columnwidth,bb= 0 0 567 459]{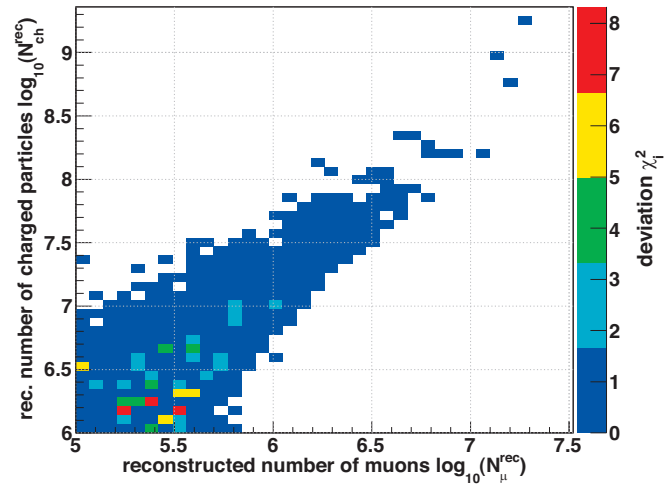}
 \caption{The distribution of the $\chi_i^2$-deviations (that are the $M$ summands $\chi_i^2$ of Eq.(\ref{eq:chi_square_test}) that contribute to the chi-square value, where $M$ is the dimension of the data vector) between the measured two-dimensional shower size plane and the one predicted by the unfolded solution.}
 \label{fig:quality_data_description_test}
\end{figure}

\section{Results}
\label{Sec:results}
In Fig.~\ref{fig:final_elemtal_spectra}, the energy spectra for elemental groups of cosmic rays, determined by applying Gold's unfolding algorithm \cite{lit:gold_alg} to the two-dimensional shower size distribution measured with KASCADE-Grande and shown in Fig.~\ref{fig:measured_shower_size_plane}, are presented. For a better distinguishability, the spectra of lighter mass groups, represented by protons as well as helium and carbon nuclei, and those of heavier mass groups, represented by silicon and iron nuclei, are depicted separately. The error bars represent the statistical uncertainties due to the limited measurement time. The error bands, representing the maximal range of systematic uncertainties, include the bias induced by Gold's algorithm, as well as the uncertainties caused by the uncertainties in the response matrix due to the limited simulation statistics. Possible uncertainties of the interaction models used, i.e. of \mbox{QGSJET-II-02} and \mbox{FLUKA} 2002.4, cannot be considered (cf. Section~\ref{Sec:error_propagation}). As emphasized in Section~\ref{Sec:monte_carlo_tests}, at the first energy bins, where full detection efficiency is not yet given, the unfolded silicon and iron spectra are subjected to larger systematic distortions. Hence, these data points are not shown. Due to the correlation among the elemental spectra, these distortions are cancelled out almost completely when computing the sum spectrum, such that it can be shown already one energy bin earlier. The intensity values of the energy spectra are listed in \ref{sec:fluxvalues}.

In the framework of the interaction models used, the cosmic ray composition is dominated by the heavy mass groups in the observed energy range. The spectra of the lighter primaries are rather structureless. There are slight indications for a recovery of protons at higher energies, which agrees with the finding in \cite{lit:KG_light_hardening} where a significant hardening in the cosmic ray spectrum of light primaries was observed. However, this is without statistical significance in this work. In case of the iron spectrum, there is a distinct knee-like steepening observed at about 80~PeV. It was verified that the spectrum of iron is not compatible with a single power law: a single power law fit results in a chi-square probability of less than 1\% ($\chi^{2}/ndf=18.9/7$). The all-particle spectrum is without significant structures. But, one has to keep in mind that the spectra of the lighter primaries suffer from larger uncertainties at higher energies, especially due to a possible strong overestimation of the fluxes caused by the positive definiteness of Gold's unfolding algorithm. Such overestimations would yield an overestimation of the sum flux at these energies, such that a possible knee structure in the all-particle spectrum at about 80~PeV to 100~PeV could maybe be masked by this effect.
\begin{figure*}[t]
 \centering
 \includegraphics[width=0.95\columnwidth,bb= 0 0 567 459]{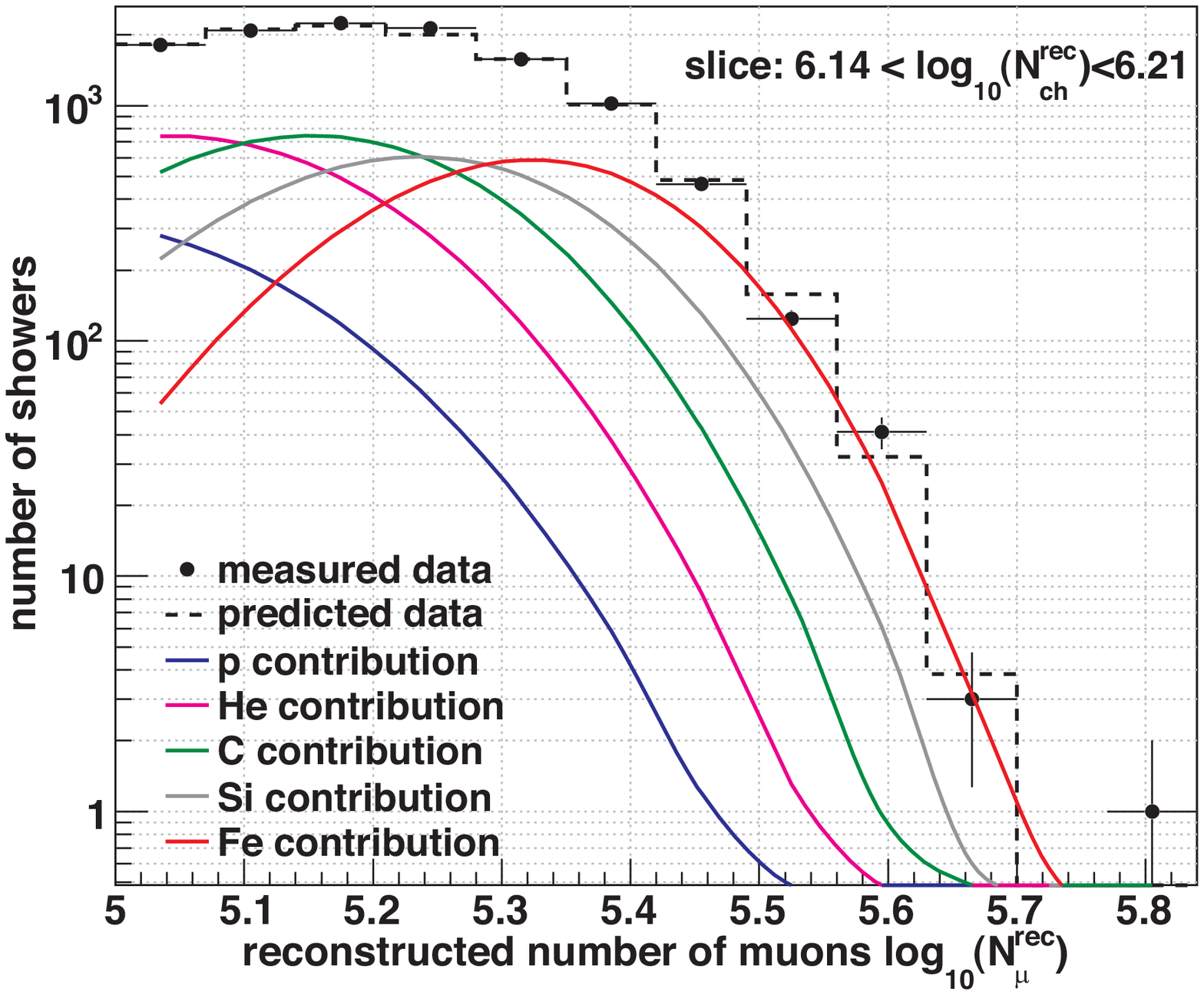}
 \includegraphics[width=0.95\columnwidth,bb= 0 0 567 459]{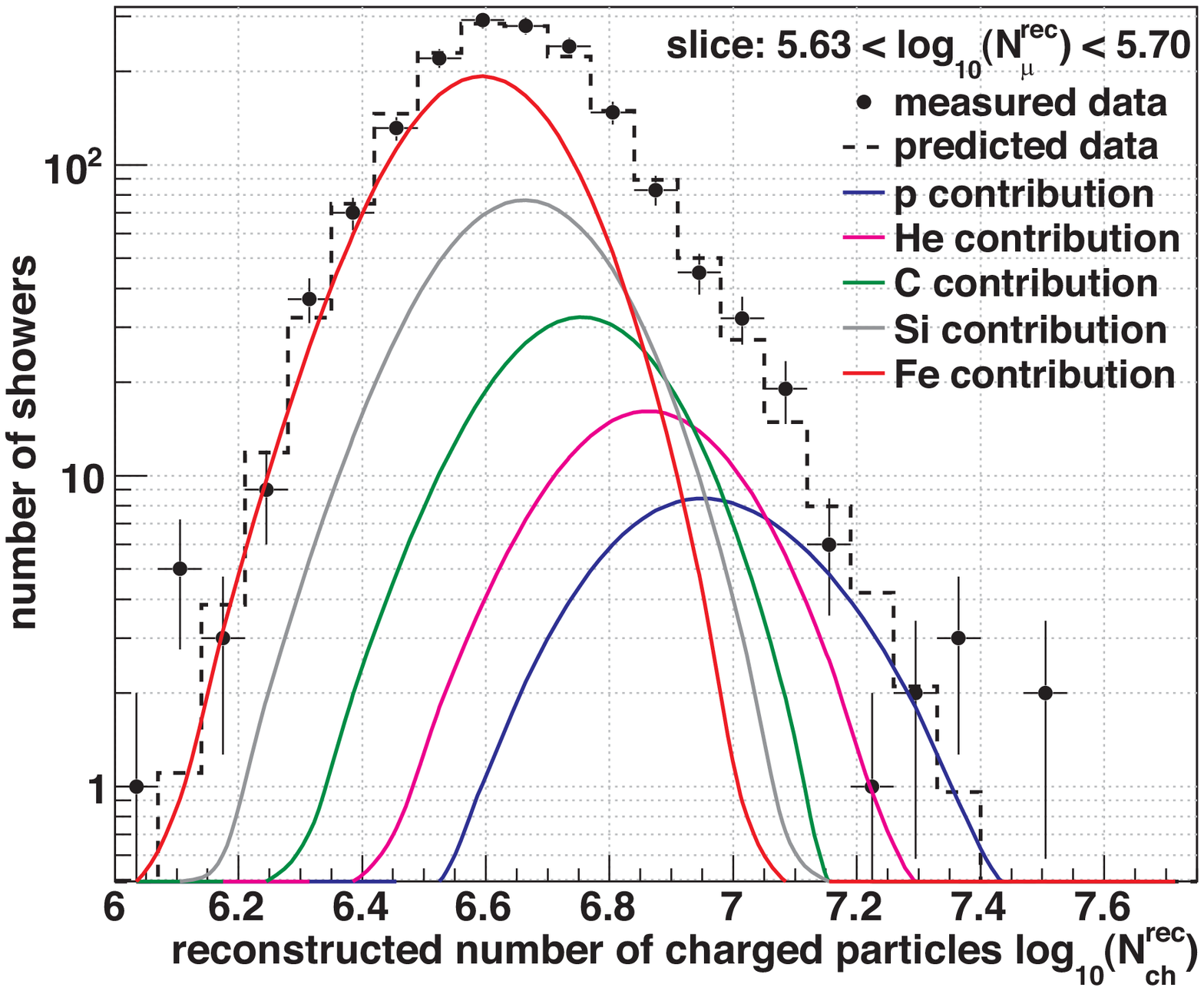}
 \caption{Slices of the two-dimensional shower size distributions, exemplarily for the fixed charged particle number interval $6.14<\log_{10}(N_{\mathrm{ch}}^{\mathrm{rec}})<6.21$ (left panel) and the muon number interval $5.63<\log_{10}(N_{\upmu}^{\mathrm{rec}})<5.70$ (right panel). The distribution predicted by the unfolded energy spectra and determined by a forward folding (dashed histogram) is compared to the distribution measured with KASCADE-Grande (markers). Additionally, the respective contributions of the individual primaries to the predicted data sample are shown (curves).}
 \label{fig:slices}
\end{figure*}

\section{Discussion of the results}
\label{Sec:discussion_results}
\subsection{Quality of data description}
\label{Sec:discussion_quality}
The quality of the unfolding solution itself cannot be judged directly, since the truth is not known. However, whether the determined energy spectra are on the main reliable can be checked indirectly by reviewing the quality of the data description by the solution. For this, the two-dimensional shower size spectrum measured with KASCADE-Grande and represented by the data vector $\overrightarrow{Y}$ can be compared to the data vector predicted by the solution vector $\overrightarrow{X}$, i.e. to the vector given by the forward folding\footnote{One has to keep in mind that a forward folding is mathematically an exact process, whereas the inversion, i.e. the unfolding procedure based on the algorithms used, is not bias free.} $\bm{R} \overrightarrow{X}$ of the solution according to Eq.(\ref{eq:faltung_matrix_equation}). If the interaction models used for the simulations are proper and if the solution is not very far from the truth, the measured data and the ones ``simulated'' by the forward folding should be in agreement. 

Firstly, a Kolmogorov-Smirnov test was applied to compare the measured two-dimensional shower size distribution with the one predicted by the solution, yielding a high probability of 97\% for a compatibility. It has to be emphasized that small deviations are possible since the measured data sample suffers from fluctuations due to the limited measurement time, while the forward folded one could be less fluctuating since a forward folding applies some kind of smoothing.

Additionally, a chi-square test is realized to compare the two-dimensional distributions: 
\begin{equation}\label{eq:chi_square_test}
\chi^2
=\dfrac{1}{M} \displaystyle \sum_{i=1}^{M}
\dfrac
{
\left(
\displaystyle \sum_{j=1}^{N}
R_{ij}x_j-y_i
 \right)^2
}
{
\sigma(y_i)^2
}
:= 
\dfrac{1}{M} \displaystyle \sum_{i=1}^{M}
\chi^2_{i}
\; \; \;.
\end{equation}
Thereby, $M$ and $N$ are the dimensions of $\overrightarrow{Y}$ and $\overrightarrow{X}$ respectively. The statistical errors $\sigma(y_i)$ of the data sample $\overrightarrow{Y}$ are assumed to be Poissonian ones, and hence are set to $\sigma(y_i)=\sqrt{y_i}$. The chi-square test results in a probability of full compatibility ($\chi^2/ndf=0.5$).
For further investigations, the distribution of the deviations $\chi^2_i$ that are the $M$ summands of Eq.(\ref{eq:chi_square_test}) is illustrated in Fig.~\ref{fig:quality_data_description_test}. Overall, there seems to be a good agreement between measurement and prediction by the unfolded solution. Only a few cells exhibit larger $\chi^2_i$ values. To examine these outliers in more detail, one-dimensional slices of the measured and of the predicted shower size plane are compared. In Fig.~\ref{fig:slices}, left panel, 
\begin{figure}[t]
 \centering
 \includegraphics[width=0.985\columnwidth,bb= 0 0 567 492]{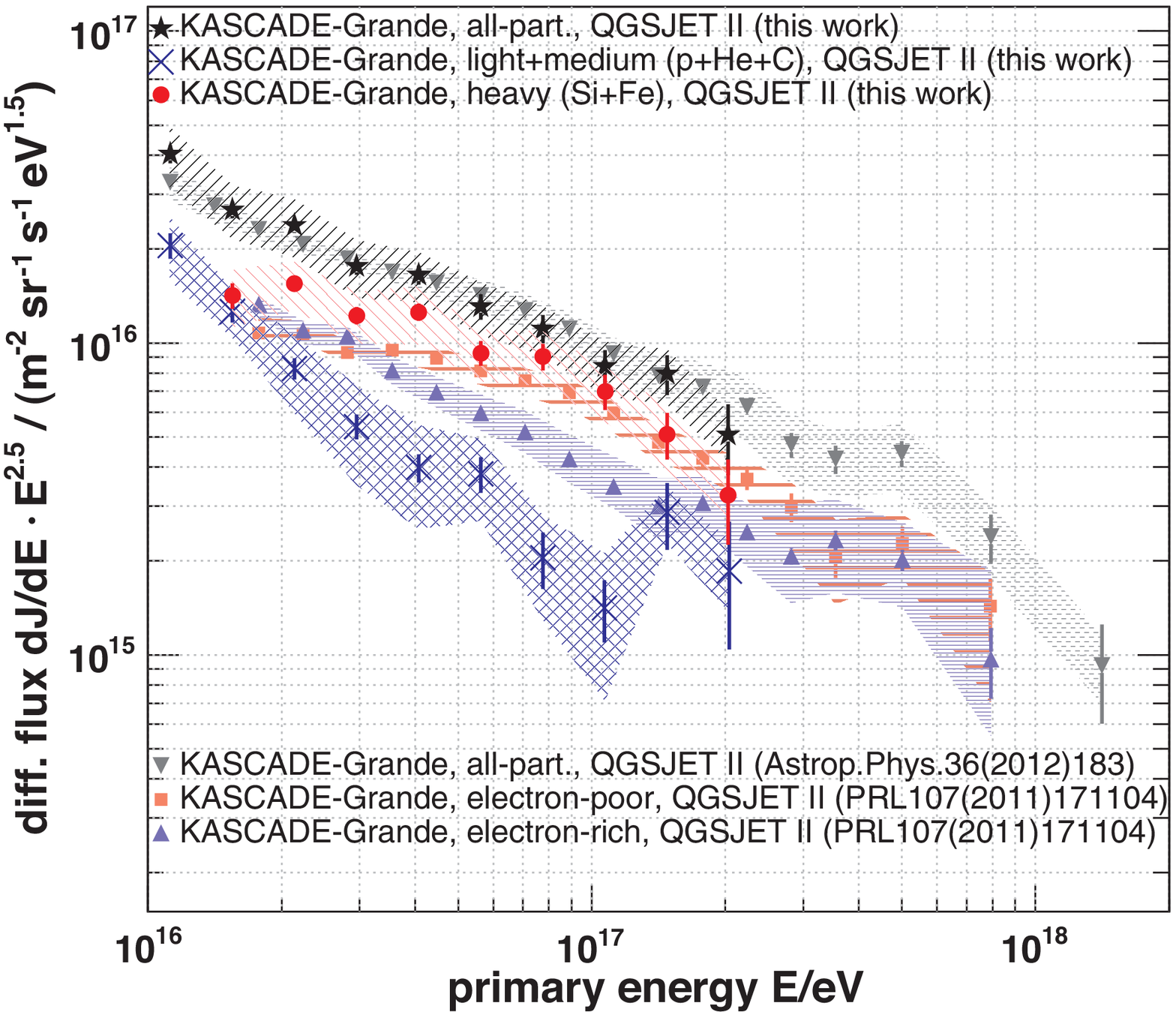}
  \caption{A comparison between the results obtained in this work with those determined with other analysis methods of the KASCADE-Grande data. The all-particle spectrum obtained in this work is compared with that shown in \cite{lit:KG_all_part_spec_paper}. Furthermore, some elemental spectra are compared, too: The electron-poor energy spectrum presented in \cite{lit:heavy_knee_paper} can roughly be compared to the sum of the elemental spectra of silicon and iron of this work (i.e. a heavy composition), and the electron-rich one to the sum of the spectra of protons, helium, and carbon (i.e. a light to intermediate composition).}
 \label{fig:comparison_KG}
\end{figure}
an exemplary slice along the x-axis of the two-dimensional shower size planes for the interval $6.14<\log_{10}(N_{\mathrm{ch}}^{\mathrm{rec}})<6.21$ is depicted. The markers represent the measured data sample, the dashed histogram the data set predicted by the forward folding of the afore unfolded solution. Additionally shown are the contributions of the considered primaries (smooth curves), determined by a forward folding of the respective elemental energy spectra into separate shower size spectra. For the examined slice there are larger deviations between measurement and prediction at the muon number bins $5.21<\log_{10}(N_{\upmu}^{\mathrm{rec}})<5.28$ and $5.49<\log_{10}(N_{\upmu}^{\mathrm{rec}})<5.56$ according to Fig.~\ref{fig:quality_data_description_test}. Following Fig.~\ref{fig:slices}, these deviations can be explained by single statistical excesses\footnote{To neglect such individual statistical excesses, only present due to the limited measurement time, is actually the goal of good unfolding algorithms.} given in the measured data sample, not present in the predicted data sample due to the smoothing property of the forward folding. There are no indications so far that the interaction models used (\mbox{QGSJET-II-02}) have problems. For the sake of completeness, slices along the y-axis are shown in Fig.~\ref{fig:slices}, right panel, again exhibiting no incompatibilities. 

To conclude, there are no indications so far that the interaction models used, i.e. \mbox{QGSJET-II-02} and \mbox{FLUKA} 2002.4, have serious deficits in the description of the physics of hadronic interactions at these energies, which, however, does not mean necessarily that these models must be accurate in all details. Different interaction models primarily have impact on the absolute scale of energy and masses, such that model uncertainties can shift the unfolded spectra, possibly resulting in different abundances of the primaries, while specific structures, e.g. knee-like features of the spectra, are less affected by the models.
\subsection{Comparison with other analyses or experiments}
\label{Sec:discussion_other_exp}
In Fig.~\ref{fig:comparison_KG}, 
the KASCADE-Grande energy spectra obtained in this work are compared with those obtained by other KASCADE-Grande analysis methods. The all-particle spectrum, which is the sum of all five unfolded elemental spectra, is compared with that shown in \cite{lit:KG_all_part_spec_paper}. Furthermore, some elemental spectra are compared, too: The electron-poor energy spectrum presented in \cite{lit:heavy_knee_paper} can roughly be compared to the sum of the elemental spectra of silicon and iron of this work (i.e. a heavy composition), and the electron-rich one to the sum of the spectra of protons, helium, and carbon (i.e. a light to intermediate composition). The sum spectra are compatible. However, in case of the elemental spectra the differences are larger, especially in the absolute flux values. The main reason is the technique used in \cite{lit:heavy_knee_paper} to divide the measured data sample into an electron-rich and electron-poor subsample in a rather simple, but robust way. Changes to the separation parameter used\footnote{In \cite{lit:heavy_knee_paper}, the separation parameter is computed by averaging the results of simulations for carbon and silicon nuclei; and hence, the transition between the two subsamples electron-poor and electron-rich takes place at a mass group between that of carbon and silicon nuclei.} affect the number of events assigned to a specific subsample, and hence affect the absolute normalization of the resulting elemental energy spectra. Thus, the differences in the absolute flux values can be interpreted by different meanings of ``light'', ``intermediate'', and ``heavy'' composition in the two compared methods, and are not an indication of inconsistencies. Despite this problem, both, the electron-poor spectrum of \cite{lit:heavy_knee_paper} and the heavy spectrum of this work, exhibit a knee-like steepening at about 80~PeV. Furthermore, both methods give slight indications for a recovery of lighter mass groups at higher energies. This is statistically not significant, but it agrees with the finding in \cite{lit:KG_light_hardening} where a significant hardening in the cosmic ray spectrum of light primaries was observed. 

\begin{figure*}[t]
 \centering
 \includegraphics[width=1.985\columnwidth,bb= 0 0 567 372]{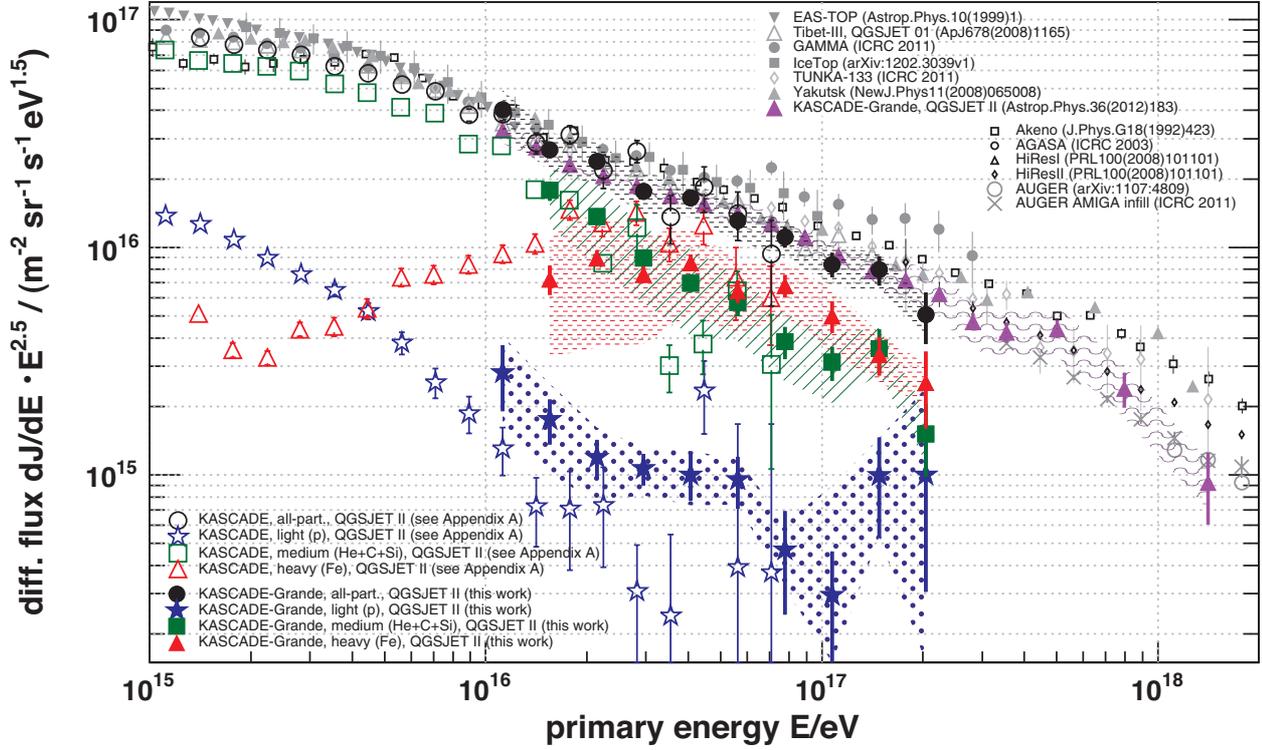}
  \caption{The all-particle spectrum obtained in this work based on an unfolding of KASCADE-Grande measurements, and the spectrum obtained in \cite{lit:marcel_phd} based on an unfolding of KASCADE measurements (see \ref{sec:finger}), are compared to spectra determined by other analysis methods of our collaboration \cite{lit:KG_all_part_spec_paper} or other experiments (see legend for references). Additionally shown are some elemental spectra representing different mass groups (see legend). The error bars denote statistical uncertainties, error bands the systematic ones (the latter ones are only shown for the results of this work, as well as for the results obtained by the alternative analysis methods of our collaboration \cite{lit:KG_all_part_spec_paper}).}
 \label{fig:comparison_all}
\end{figure*}

Figure~\ref{fig:comparison_all} compiles
the KASCADE-Grande energy spectra obtained in this work with those obtained by other KASCADE-Grande analysis methods or by other experiments. For a better distinguishability, the three intermediate spectra of helium, carbon, and silicon are summed up to one ``medium spectrum''. Within the given uncertainties, the KASCADE-Grande all-particle spectra are compatible with those of most of the other experiments. Though, at higher energies, the KASCADE-Grande spectrum exhibits a lower intensity compared to earlier experiments, especially GAMMA, Akeno and Yakutsk. At the highest energies, the KASCADE-Grande statistics are low and the all-particle spectrum is compatible with a single power law. Assuming a single power law fit, our results are in agreement with those reported by HiRes and the Pierre Auger Observatory. Concerning the elemental energy spectra of different mass groups, there is a good agreement with the new \mbox{QGSJET-II-02} based results~\cite{lit:marcel_phd} of the KASCADE experiment, despite the independent measurement and data analysis. A brief discussion on this KASCADE analysis is compiled in~\ref{sec:finger}.
Combining the findings of KASCADE and KASCADE-Grande, all elemental spectra exhibit a knee-like structure: for light primaries at about 3~PeV to 5~PeV, for medium ones at about 8~PeV to 10~PeV, and for heavy ones at about 80~PeV.
\section{Summary and conclusion}
\label{Sec:summary}
The two-dimensional shower size spectrum of charged particles and muons measured with KASCADE-Grande was unfolded. Based on this analysis, the energy spectra for five primaries representing the chemical composition of cosmic rays have been determined, as well as the all-particle spectrum which is the sum of the elemental spectra. For this analysis, the response matrix of the experiment was computed based on the hadronic interaction models \mbox{QGSJET-II-02}~\citep{lit:qgsjet-ii-model-zitat1, lit:qgsjet-ii-model-zitat2} and \mbox{FLUKA} 2002.4~\citep{lit:FlukaAllgemeinZitat1,lit:Fluka2002_4_Zitat1,lit:FlukaAllgemeinZitat2}.

The all-particle spectrum, which suffers in this work from uncertainties of the contributing elemental spectra and which is structureless within the given uncertainties, agrees with that determined in an alternative analysis of the KASCADE-Grande data \cite{lit:KG_all_part_spec_paper}, where a small break-off at about 80~PeV was found\footnote{In the energy range from 1~PeV to some hundred PeV, this break-off in the all-particle spectrum is the second one besides the one at about 3~PeV to 5~PeV reported in \cite{lit:marcel_phd} based on KASCADE data an using also \mbox{QGSJET-II-02} as interaction model.}. Furthermore, both KASCADE-Grande all-particle spectra are compatible with the findings of most of the other experiments. 

The unfolded energy spectra of light and intermediate primaries are rather featureless in the sensitive energy range. There are slight indications for a possible recovery of protons at higher energies, which is, however, statistically not significant. But, this finding would agree with the one in \cite{lit:KG_light_hardening} where a significant hardening in the cosmic ray spectrum of light primaries was observed.

The spectrum of iron exhibits a clear knee-like structure at about 80~PeV. The position of this structure is consistent with that of a structure found in spectra of heavy primaries determined by other analysis methods of the KASCADE-Grande data \cite{lit:heavy_knee_paper}. The energy where this knee-like structure occurs conforms to the one where the break-off in the all-particle spectrum is observed. Hence, the findings in this work and in \cite{lit:heavy_knee_paper} demonstrate the first time experimentally that the heavy knee exists, and the kink in the all-particle spectrum is presumably caused by this decrease in the flux of heavy primaries. The spectral steepening occurs at an energy where the charge dependent knee of iron is expected, if the knee at about 3~PeV to 5~PeV is assumed to be caused by a decrease in the flux of light primaries (protons and/or helium). 

However, there is still uncertainty about whether the applied interaction models, especially the high energy one \mbox{QGSJET-II-02}, are valid in all the details. As demonstrated in \cite{lit:kascade-unfolding}, it is expected that variations in the interaction models primarily affect the relative abundances of the primaries, and hence assign possible structures given in the data to different mass groups, while the structures themselves are rather model independent. Although it was shown that the interaction models used do not seem to exhibit significant weaknesses in describing the data, more certainty can be expected in the near future, when man-made particle accelerators like the LHC reach laboratory energies up to some hundred PeV, and hence allow to optimize the interaction models in an energy range relevant for KASCADE-Grande.
\section*{Acknowledgements}
\label{Sec:Acknowledgments}
The authors would like to thank the members of the
engineering and technical staff of the KASCADE-Grande
collaboration, who contributed to the success of the experiment.
The KASCADE-Grande experiment is supported
by the BMBF of Germany, the MIUR and INAF of Italy,
the Polish Ministry of Science and Higher Education,
and the Romanian Authority for Scientific Research UEFISCDI 
(PNII-IDEI grants 271/2011 and 17/2011).






\appendix

\section{KASCADE data unfolding based on QGSJET-II}
\label{sec:finger}
\begin{figure*}[t]
 \centering
 \includegraphics[width=0.95\columnwidth]{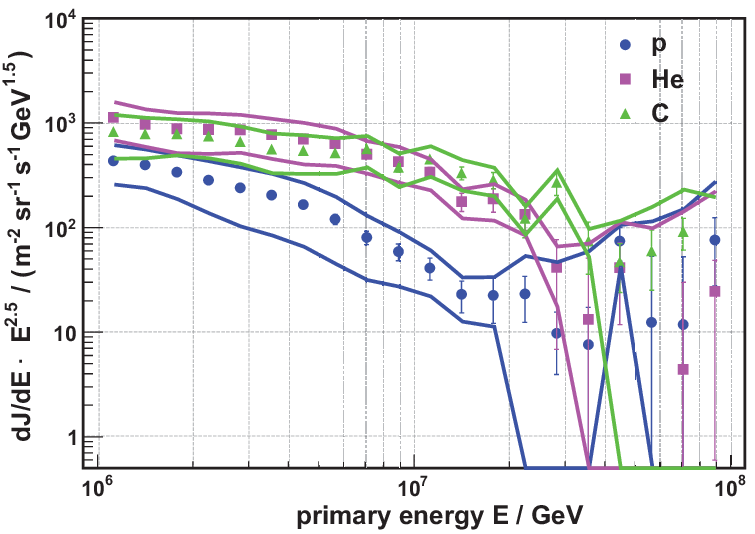}
 \includegraphics[width=0.95\columnwidth]{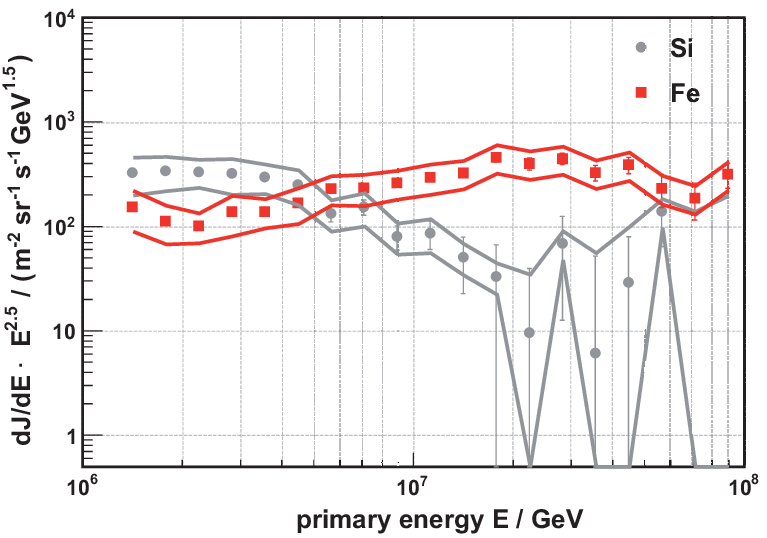}
 \caption{Unfolded energy spectra of protons as well as helium and carbon nuclei (left panel), and silicon and iron nuclei (right panel), using as hadronic interaction models \mbox{QGSJET-II-02}~\citep{lit:qgsjet-ii-model-zitat1, lit:qgsjet-ii-model-zitat2} and \mbox{FLUKA} 2002.4~\citep{lit:FlukaAllgemeinZitat1,lit:Fluka2002_4_Zitat1,lit:FlukaAllgemeinZitat2}. The error bands mark the maximal range of systematic uncertainties, and the error bars represent the statistical uncertainties. Basis for this analysis are air showers measured with the KASCADE experiment for zenith angles from $0^\circ$ to $18^\circ$.}
 \label{fig:specfinger}
\end{figure*}
In Fig.~\ref{fig:comparison_all}, the results obtained by an unfolding analysis applied to air showers measured with the KASCADE experiment \cite{lit:kascade_allgemein_nimpaper} in the zenith angle range of  
$0^\circ$ to $18^\circ$ are depicted.  
In this appendix, we will discuss briefly the main findings of the corresponding analysis, 
while details can be found in~\cite{lit:marcel_phd}.
The analysis is based on the same method of data unfolding and the same hadronic interaction models (\mbox{QGSJET-II-02}~\citep{lit:qgsjet-ii-model-zitat1, lit:qgsjet-ii-model-zitat2} and \mbox{FLUKA} 2002.4~\citep{lit:FlukaAllgemeinZitat1,lit:Fluka2002_4_Zitat1,lit:FlukaAllgemeinZitat2}) as the work 
described in this paper. But, instead of the total number of charged particles $N_{\mathrm{ch}}$ and the total number of muons $N_{\upmu}$, here the electron shower size $N_{\mathrm{e}}$ and the truncated muon number $N_{\upmu}^{\mathrm{trunc}}$ (number of muons between $40$~m and $200$~m distance from shower core) are used. 
Another difference is that KASCADE covers a lower energy range than KASCADE-Grande, but a reasonable overlap remains. 

At an energy of approximately 4~PeV to 5~PeV, a kink in the all-particle flux, the so-called ``knee'' of the cosmic ray spectrum, can be observed (cf. Fig.~\ref{fig:comparison_all}).
The left panel of Fig.~\ref{fig:specfinger} shows the energy spectra of protons, as well as of helium and carbon nuclei. It 
can be noticed that, in the frame of the models used, protons are less abundant than helium and carbon nuclei, which is in agreement with the results at higher energies (cf. Fig.~\ref{fig:final_elemtal_spectra}). 
At an energy of about 4~PeV, a kink 
in the proton spectrum can be found. The energy spectra of helium and carbon, which
are the most abundant nuclei, indicate an almost equal abundance of both elements, but the fluxes of the two primary particle types differ in their spectral shape. 
Whereas the helium spectrum is characterized by a kink at about 7~PeV, 
a change of index in the carbon spectrum is compatible with a kink at around 20~PeV. As discussed in~\cite{lit:kascade-unfolding} for other models, the knee positions of the three nuclei protons, helium, and carbon relative to each other demonstrate a compatibility with a rigidity
dependence of the knees. It should be mentioned that in case of the
steepening of the carbon spectrum the statistics become poor in this energy region and the
spectrum is liable to large fluctuations; but, a general trend can be seen.
The right part of Fig.~\ref{fig:specfinger} exhibits the energy spectra of silicon and of iron nuclei. The silicon
spectrum reveals a kink at quite low energy, which is not expected when a rigidity dependence is assumed. 
Its existence can be explained by problems in the data description. 
An examination of the distribution of the $\chi_i^2$-deviations (analogous to the examination performed in Section~\ref{Sec:discussion_quality} for the KASCADE-Grande data) reveals deficiencies
mainly in the medium energy range, especially at the heavy ridge\footnote{The definition of light and heavy ridge is given in Fig.~\ref{fig:measured_shower_size_plane} and its caption.}, which might explain the unexpected course of the silicon spectrum. 
But, in general, the distribution of the $\chi_i^2$-deviations indicates a good overall data description compared to other models.
The spectrum of iron does not exhibit a knee-like feature in the accessible energy range of KASCADE.  

\begin{figure}[t]
 \centering
 \includegraphics[width=0.985\columnwidth,bb=0 0 252 158]{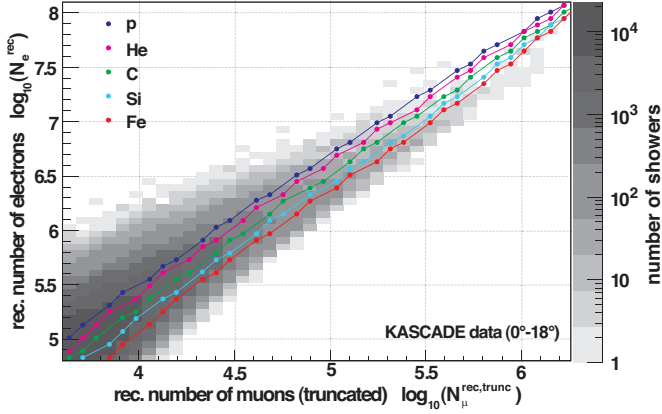}
  \caption{Most probable values for the five considered nuclei according to the calculated response matrix (based on the interaction models \mbox{QGSJET-II-02}~\citep{lit:qgsjet-ii-model-zitat1, lit:qgsjet-ii-model-zitat2} and \mbox{FLUKA} 2002.4~\citep{lit:FlukaAllgemeinZitat1,lit:Fluka2002_4_Zitat1,lit:FlukaAllgemeinZitat2}). Additionally, the measured two-dimensional size spectrum (the
data used) is depicted.}
 \label{fig:mopfinger}
\end{figure}

Figure~\ref{fig:mopfinger} shows the measured two-dimensional shower size spectrum of electron and muon numbers and,
additionally, the lines of the most probable values (given by the response matrix) for all nuclei used. Whereas the lines of
helium and carbon, being the most abundant elements at low energies, start almost in the maximum
of the measured size spectrum, the line of protons is located on the left-hand side
of the maximum, which causes a minor frequency of protons. With increasing energy, the most
probable values leave, for all primaries consecutively, the maximum region, leading to a kink in the individual
energy spectra. For silicon, the situation seems to be more complicated. Although the lines start
on the right-hand side of the maximum of the data distribution and converge with increasing
energy towards the maximum, a sharp kink in the silicon spectrum can be found. Solely from
an examination of the course of the most probable values, this change of index is not expected.
With respect to the most probable values of silicon, the values for iron are shifted to the heavy
edge. The deficit of iron at low energies as well as the kink in the silicon spectrum cannot be
clarified by analysing the most probable values only. This reveals the importance of the shower
fluctuations for the results of the analysis.

In summary, the presented KASCADE unfolding analysis, which is based on the high energy interaction model \mbox{QGSJET-II-02}~\citep{lit:qgsjet-ii-model-zitat1, lit:qgsjet-ii-model-zitat2} and the low energy interaction model \mbox{FLUKA} 2002.4~\citep{lit:FlukaAllgemeinZitat1,lit:Fluka2002_4_Zitat1,lit:FlukaAllgemeinZitat2}, confirms especially the former finding that the knee in the cosmic ray energy spectrum is caused by a decrease of the flux of the lighter mass groups, like already shown in \cite{lit:kascade-unfolding}, where the KASCADE data was unfolded \textit{inter alia} based on the high energy interaction model \mbox{QGSJET 01}~\cite{lit:qgsjet01} and the low energy interaction model \mbox{GHEISHA}~\cite{lit:GHEISHA}, version 2002. The influence of the low energy interaction model on the resulting spectra is small compared to the systematic uncertainties (shown and discussed in \cite{lit:KASCADE_low_energy_interaction_model}).
But, as found in \cite{lit:kascade-unfolding}, the choice of the high energy interaction model affects the relative abundance of the primary mass groups, though not the spectral shapes. As for the KASCADE-Grande analysis the more sophisticated \mbox{QGSJET-II-02} model was used instead of \mbox{QGSJET 01}, we repeated the analysis of the KASCADE data based on \mbox{QGSJET-II-02}. Figure~\ref{fig:specfinger} can be compared directly with Fig.~14 of reference \cite{lit:kascade-unfolding}: whereas the structural features of the light components are similar, the spectrum of the iron component changed. This is due to the improved description of the shower development by the new version of the model. Also the overall description of the data (in terms of $\chi_i^2$-deviations) is improved considerably.
\section{Differential intensity values}
\label{sec:fluxvalues}

The differential intensity values d$J$/d$E$ of the unfolded energy spectra for elemental groups of cosmic rays---based on KASCADE-Grande measurements and depicted in Fig.~\ref{fig:final_elemtal_spectra}---and their statistical and systematic uncertainties, $\sigma_{\mathrm{stat.}}$ respectively $\Delta_{\mathrm{syst.}}$, are listed in Table~\ref{tab:flux_values_final_result_H} to \ref{tab:flux_values_final_result_tot}. The results are based on the interaction models \mbox{QGSJET-II-02}~\citep{lit:qgsjet-ii-model-zitat1, lit:qgsjet-ii-model-zitat2} and \mbox{FLUKA} 2002.4~\citep{lit:FlukaAllgemeinZitat1,lit:Fluka2002_4_Zitat1,lit:FlukaAllgemeinZitat2}.
\begin{table}[ht]
 \centering
 \begin{tabular}[t]{c||c}
 Energy $\dfrac{E}{\mathrm{GeV}}$    &    $\dfrac{\mathrm{d}J/\mathrm{d}E \pm \sigma_{\mathrm{stat.}} \pm \Delta_{\mathrm{syst.}}} {\mathrm m^{-2} \mathrm{sr}^{-1} \mathrm{s}^{-1} \mathrm{GeV}^{-1}}$   \\ \hline \hline
 1.12$\times 10^{7}$ & (2.11$\:\pm\:$0.68$\:\pm\:$0.93)$\times 10^{-16}$ \\ \hline
 1.55$\times 10^{7}$ & (5.87$\:\pm\:$1.31$\:\pm\:$2.54)$\times 10^{-17}$ \\ \hline
 2.14$\times 10^{7}$ & (1.78$\:\pm\:$0.35$\:\pm\:$0.63)$\times 10^{-17}$ \\ \hline
 2.95$\times 10^{7}$ & (7.13$\:\pm\:$1.12$\:\pm\:$1.69)$\times 10^{-18}$ \\ \hline
 4.07$\times 10^{7}$ & (2.99$\:\pm\:$0.69$\:\pm\:$0.80)$\times 10^{-18}$ \\ \hline
 5.62$\times 10^{7}$ & (1.27$\:\pm\:$0.33$\:\pm\:$0.29)$\times 10^{-18}$ \\ \hline
 7.76$\times 10^{7}$ & (2.79$\:\pm\:$1.34$\:\pm\:$0.89)$\times 10^{-19}$ \\ \hline
 1.07$\times 10^{8}$ & (7.87$\:\pm\:$4.41$\:\pm\:$0.16)$\times 10^{-20}$ \\ \hline
 1.48$\times 10^{8}$ & (1.18$\:\pm\:$0.55$\:\pm\:$0.54)$\times 10^{-19}$ \\ \hline
 2.04$\times 10^{8}$ & (5.30$\:\pm\:$3.68$\:\pm\:$7.49)$\times 10^{-20}$ 
 \end{tabular}
\caption{Differential flux values d$J$/d$E$ and uncertainties $\sigma_{\mathrm{stat.}}$ respectively $\Delta_{\mathrm{syst.}}$ for protons. The response matrix used bases on the interaction models \mbox{QGSJET-II-02}~\citep{lit:qgsjet-ii-model-zitat1, lit:qgsjet-ii-model-zitat2} and \mbox{FLUKA} 2002.4~\citep{lit:FlukaAllgemeinZitat1,lit:Fluka2002_4_Zitat1,lit:FlukaAllgemeinZitat2}.}
 \label{tab:flux_values_final_result_H}
\end{table}
\begin{table}[ht]
 \centering
 \begin{tabular}[t]{c||c}
 Energy $\dfrac{E}{\mathrm{GeV}}$    &    $\dfrac{\mathrm{d}J/\mathrm{d}E \pm \sigma_{\mathrm{stat.}} \pm \Delta_{\mathrm{syst.}}} {\mathrm m^{-2} \mathrm{sr}^{-1} \mathrm{s}^{-1} \mathrm{GeV}^{-1}}$   \\ \hline \hline
1.12$\times 10^{7}$ & (5.75$\:\pm\:$0.72$\:\pm\:$1.98)$\times 10^{-16}$ \\ \hline
1.55$\times 10^{7}$ & (1.43$\:\pm\:$0.19$\:\pm\:$0.63)$\times 10^{-16}$ \\ \hline
2.14$\times 10^{7}$ & (3.72$\:\pm\:$0.65$\:\pm\:$2.02)$\times 10^{-17}$ \\ \hline
2.95$\times 10^{7}$ & (1.05$\:\pm\:$0.20$\:\pm\:$0.62)$\times 10^{-17}$ \\ \hline
4.07$\times 10^{7}$ & (3.28$\:\pm\:$0.58$\:\pm\:$1.98)$\times 10^{-18}$ \\ \hline
5.62$\times 10^{7}$ & (1.62$\:\pm\:$0.36$\:\pm\:$0.76)$\times 10^{-18}$ \\ \hline
7.76$\times 10^{7}$ & (3.55$\:\pm\:$1.12$\:\pm\:$1.76)$\times 10^{-19}$ \\ \hline
1.07$\times 10^{8}$ & (1.03$\:\pm\:$0.42$\:\pm\:$1.85)$\times 10^{-19}$ \\ \hline
1.48$\times 10^{8}$ & (1.11$\:\pm\:$0.35$\:\pm\:$0.42)$\times 10^{-19}$ \\ \hline
2.04$\times 10^{8}$ & (2.22$\:\pm\:$1.19$\:\pm\:$8.04)$\times 10^{-20}$ 
\end{tabular}
\caption{Differential flux values d$J$/d$E$ and uncertainties $\sigma_{\mathrm{stat.}}$ respectively $\Delta_{\mathrm{syst.}}$ for helium nuclei. The response matrix used bases on the interaction models \mbox{QGSJET-II-02}~\citep{lit:qgsjet-ii-model-zitat1, lit:qgsjet-ii-model-zitat2} and \mbox{FLUKA} 2002.4~\citep{lit:FlukaAllgemeinZitat1,lit:Fluka2002_4_Zitat1,lit:FlukaAllgemeinZitat2}.}
 \label{tab:flux_values_final_result_He}
\end{table}
\begin{table}[ht]
 \centering
 \begin{tabular}[t]{c||c}
 Energy $\dfrac{E}{\mathrm{GeV}}$    &    $\dfrac{\mathrm{d}J/\mathrm{d}E \pm \sigma_{\mathrm{stat.}} \pm \Delta_{\mathrm{syst.}}} {\mathrm m^{-2} \mathrm{sr}^{-1} \mathrm{s}^{-1} \mathrm{GeV}^{-1}}$   \\ \hline \hline
 1.12$\times 10^{7}$ & (7.57$\:\pm\:$1.03$\:\pm\:$1.92)$\times 10^{-16}$ \\ \hline
 1.55$\times 10^{7}$ & (2.20$\:\pm\:$0.24$\:\pm\:$0.88)$\times 10^{-16}$ \\ \hline
 2.14$\times 10^{7}$ & (6.91$\:\pm\:$0.67$\:\pm\:$1.49)$\times 10^{-17}$ \\ \hline
 2.95$\times 10^{7}$ & (1.85$\:\pm\:$0.23$\:\pm\:$0.57)$\times 10^{-17}$ \\ \hline
 4.07$\times 10^{7}$ & (5.58$\:\pm\:$0.86$\:\pm\:$2.12)$\times 10^{-18}$ \\ \hline
 5.62$\times 10^{7}$ & (2.19$\:\pm\:$0.45$\:\pm\:$0.59)$\times 10^{-18}$ \\ \hline
 7.76$\times 10^{7}$ & (5.91$\:\pm\:$1.82$\:\pm\:$1.80)$\times 10^{-19}$ \\ \hline
 1.07$\times 10^{8}$ & (1.94$\:\pm\:$0.60$\:\pm\:$0.89)$\times 10^{-19}$ \\ \hline
 1.48$\times 10^{8}$ & (1.11$\:\pm\:$0.50$\:\pm\:$0.42)$\times 10^{-19}$ \\ \hline
 2.04$\times 10^{8}$ & (2.34$\:\pm\:$1.95$\:\pm\:$7.31)$\times 10^{-20}$ 
\end{tabular}
\caption{Differential flux values d$J$/d$E$ and uncertainties $\sigma_{\mathrm{stat.}}$ respectively $\Delta_{\mathrm{syst.}}$ for carbon nuclei. The response matrix used bases on the interaction models \mbox{QGSJET-II-02}~\citep{lit:qgsjet-ii-model-zitat1, lit:qgsjet-ii-model-zitat2} and \mbox{FLUKA} 2002.4~\citep{lit:FlukaAllgemeinZitat1,lit:Fluka2002_4_Zitat1,lit:FlukaAllgemeinZitat2}.}
 \label{tab:flux_values_final_result_C}
\end{table}
\begin{table}[ht]
 \centering
 \begin{tabular}[t]{c||c}
 Energy $\dfrac{E}{\mathrm{GeV}}$    &    $\dfrac{\mathrm{d}J/\mathrm{d}E \pm \sigma_{\mathrm{stat.}} \pm \Delta_{\mathrm{syst.}}} {\mathrm m^{-2} \mathrm{sr}^{-1} \mathrm{s}^{-1} \mathrm{GeV}^{-1}}$   \\ \hline \hline
 1.55$\times 10^{7}$ & (2.32$\:\pm\:$0.27$\:\pm\:$1.12)$\times 10^{-16}$ \\ \hline
 2.14$\times 10^{7}$ & (9.79$\:\pm\:$0.80$\:\pm\:$7.30)$\times 10^{-17}$ \\ \hline
 2.95$\times 10^{7}$ & (3.10$\:\pm\:$0.28$\:\pm\:$2.41)$\times 10^{-17}$ \\ \hline
 4.07$\times 10^{7}$ & (1.19$\:\pm\:$0.11$\:\pm\:$0.97)$\times 10^{-17}$ \\ \hline
 5.62$\times 10^{7}$ & (3.83$\:\pm\:$0.52$\:\pm\:$2.54)$\times 10^{-18}$ \\ \hline
 7.76$\times 10^{7}$ & (1.35$\:\pm\:$0.26$\:\pm\:$0.79)$\times 10^{-18}$ \\ \hline
 1.07$\times 10^{8}$ & (5.35$\:\pm\:$1.14$\:\pm\:$3.33)$\times 10^{-19}$ \\ \hline
 1.48$\times 10^{8}$ & (2.04$\:\pm\:$0.70$\:\pm\:$1.07)$\times 10^{-19}$ \\ \hline
 2.04$\times 10^{8}$ & (3.49$\:\pm\:$1.67$\:\pm\:$7.33)$\times 10^{-20}$ 
 \end{tabular}
\caption{Differential flux values d$J$/d$E$ and uncertainties $\sigma_{\mathrm{stat.}}$ respectively $\Delta_{\mathrm{syst.}}$ for silicon nuclei. The response matrix used bases on the interaction models \mbox{QGSJET-II-02}~\citep{lit:qgsjet-ii-model-zitat1, lit:qgsjet-ii-model-zitat2} and \mbox{FLUKA} 2002.4~\citep{lit:FlukaAllgemeinZitat1,lit:Fluka2002_4_Zitat1,lit:FlukaAllgemeinZitat2}.}
 \label{tab:flux_values_final_result_Si}
\end{table}
\begin{table}[ht]
 \centering
 \begin{tabular}[t]{c||c}
 Energy $\dfrac{E}{\mathrm{GeV}}$    &    $\dfrac{\mathrm{d}J/\mathrm{d}E \pm \sigma_{\mathrm{stat.}} \pm \Delta_{\mathrm{syst.}}} {\mathrm m^{-2} \mathrm{sr}^{-1} \mathrm{s}^{-1} \mathrm{GeV}^{-1}}$   \\ \hline \hline
 1.55$\times 10^{7}$ & (2.43$\:\pm\:$0.35$\:\pm\:$1.33)$\times 10^{-16}$ \\ \hline
 2.14$\times 10^{7}$ & (1.34$\:\pm\:$0.11$\:\pm\:$0.80)$\times 10^{-16}$ \\ \hline
 2.95$\times 10^{7}$ & (5.09$\:\pm\:$0.42$\:\pm\:$2.58)$\times 10^{-17}$ \\ \hline
 4.07$\times 10^{7}$ & (2.56$\:\pm\:$0.21$\:\pm\:$1.03)$\times 10^{-17}$ \\ \hline
 5.62$\times 10^{7}$ & (8.58$\:\pm\:$1.00$\:\pm\:$3.07)$\times 10^{-18}$ \\ \hline
 7.76$\times 10^{7}$ & (4.05$\:\pm\:$0.46$\:\pm\:$1.08)$\times 10^{-18}$ \\ \hline
 1.07$\times 10^{8}$ & (1.33$\:\pm\:$0.21$\:\pm\:$0.41)$\times 10^{-18}$ \\ \hline
 1.48$\times 10^{8}$ & (4.01$\:\pm\:$0.77$\:\pm\:$0.99)$\times 10^{-19}$ \\ \hline
 2.04$\times 10^{8}$ & (1.35$\:\pm\:$0.50$\:\pm\:$0.28)$\times 10^{-19}$ 
 \end{tabular}
\caption{Differential flux values d$J$/d$E$ and uncertainties $\sigma_{\mathrm{stat.}}$ respectively $\Delta_{\mathrm{syst.}}$ for iron nuclei. The response matrix used bases on the interaction models \mbox{QGSJET-II-02}~\citep{lit:qgsjet-ii-model-zitat1, lit:qgsjet-ii-model-zitat2} and \mbox{FLUKA} 2002.4~\citep{lit:FlukaAllgemeinZitat1,lit:Fluka2002_4_Zitat1,lit:FlukaAllgemeinZitat2}.}
\end{table}
\begin{table}[ht]
 \centering
 \begin{tabular}[t]{c||c|c}
 Energy $\dfrac{E}{\mathrm{GeV}}$    &    $\dfrac{\mathrm{d}J/\mathrm{d}E \pm \sigma_{\mathrm{stat.}} \pm \Delta_{\mathrm{syst.}}} {\mathrm m^{-2} \mathrm{sr}^{-1} \mathrm{s}^{-1} \mathrm{GeV}^{-1}}$  &  $N$  \\ \hline \hline
 1.12$\times 10^{7}$ & (3.03$\:\pm\:$0.20$\:\pm\:$0.66)$\times 10^{-15}$ & $56\,900$ \\ \hline
 1.55$\times 10^{7}$ & (8.97$\:\pm\:$0.55$\:\pm\:$1.72)$\times 10^{-16}$ & $23\,300$ \\ \hline
 2.14$\times 10^{7}$ & (3.57$\:\pm\:$0.19$\:\pm\:$0.74)$\times 10^{-16}$ & $12\,800$ \\ \hline
 2.95$\times 10^{7}$ & (1.18$\:\pm\:$0.07$\:\pm\:$0.25)$\times 10^{-16}$ & $5\,830$ \\ \hline
 4.07$\times 10^{7}$ & (4.94$\:\pm\:$0.34$\:\pm\:$0.89)$\times 10^{-17}$ & $3\,370$ \\ \hline
 5.62$\times 10^{7}$ & (1.75$\:\pm\:$0.16$\:\pm\:$0.33)$\times 10^{-17}$ & $1\,650$ \\ \hline
 7.76$\times 10^{7}$ & (6.62$\:\pm\:$0.69$\:\pm\:$1.18)$\times 10^{-18}$ & $861$ \\ \hline
 1.07$\times 10^{8}$ & (2.24$\:\pm\:$0.28$\:\pm\:$0.44)$\times 10^{-18}$ & $402$ \\ \hline
 1.48$\times 10^{8}$ & (9.45$\:\pm\:$1.37$\:\pm\:$1.86)$\times 10^{-19}$ & $234$ \\ \hline
 2.04$\times 10^{8}$ & (2.68$\:\pm\:$0.69$\:\pm\:$0.48)$\times 10^{-19}$ & $92$
\end{tabular}
\caption{Differential flux values d$J$/d$E$ and uncertainties $\sigma_{\mathrm{stat.}}$ respectively $\Delta_{\mathrm{syst.}}$, as well as the absolute number of events $N$ for the all-particle spectrum. The response matrix used bases on the interaction models \mbox{QGSJET-II-02}~\citep{lit:qgsjet-ii-model-zitat1, lit:qgsjet-ii-model-zitat2} and \mbox{FLUKA} 2002.4~\citep{lit:FlukaAllgemeinZitat1,lit:Fluka2002_4_Zitat1,lit:FlukaAllgemeinZitat2}.}
 \label{tab:flux_values_final_result_tot}
\end{table}

\end{document}